\documentclass[aip,jap,reprint,showkeys]{revtex4-1}

\usepackage{color}
\usepackage{graphicx}
\usepackage{amssymb}
\usepackage{amsmath}
\usepackage{textcomp}
\usepackage{epstopdf}

\begin{document}

%\preprint{AIP/123-QED}

\title{Second harmonic microscopy of poled x-cut thin film lithium niobate: Understanding the contrast mechanism}

\author{M. R\"using}
 \email{mruesing@eng.ucsd.edu}
\affiliation{Department of Electrical and Computer Engineering, University of California, San Diego, La Jolla, CA-92093-0407, USA}

\author{J. Zhao}
\affiliation{Department of Electrical and Computer Engineering, University of California, San Diego, La Jolla, CA-92093-0407, USA}

\author{S. Mookherjea}
\affiliation{Department of Electrical and Computer Engineering, University of California, San Diego, La Jolla, CA-92093-0407, USA}

\date{\today}% It is always \today, today,

\begin{abstract}
Thin film lithium niobate is of great recent interest and an understanding of periodically poled thin-films is crucial for both fundamental physics and device developments. Second-harmonic (SH) microscopy allows for the non-invasive visualization and analysis of ferroelectric domain structures and walls. While the technique is well understood in bulk lithium niobate, SH microscopy in thin films is largely influenced by interfacial reflections and resonant enhancements, which depend on film thicknesses and the substrate materials. We present a comprehensive analysis of SH microscopy in x-cut lithium niobate thin films, based on a full three dimensional focus calculations, and accounting for interface reflections. We show that the dominant signal in back-reflection originates from a co-propagating phase-matched process observed through reflections, rather than direct detection of the counter-propagating signal as in bulk samples. We can explain the observation of domain structures in the thin film geometry, and in particular, we show that the SH signal from thin poled films allows to unambiguously distinguish areas, which are completely or only partly inverted in depth. 
\end{abstract}

%Use showkeys class option if keyword display desired
\maketitle

\section{\label{sec:Intro}Introduction}

Second-harmonic (SH) microscopy allows the non-invasive and non-destructive visualization and analysis of ferroelectric domain structures and crystal properties, and requires no sample preparation  \cite{Berth2007,Berth2009,Mackwitz2016,Florsheimer1998,Bozhevolnyi1998,Cherifi-Hertel2017,Huang2017,Zhao2019,Kaneshiro2010,Kurimura1997}. It makes use of the fact that SH generation is very sensitive to local changes in symmetry or crystal structure \cite{Cherifi-Hertel2017,Spychala2017}. The interest in tailored ferroelectric domain structures ranges from fundamental physics, e.g. for the study of topologically protected properties \cite{Wu2012,Kumagai2013,Ma2018,Nahas2015} and to potentially novel type of electronics in conductive domain walls (DWs) \cite{Sharma2017,Godau2017,Schroder2012,Wolba2018}, to commercialized applications, such as in nonlinear optical frequency converters \cite{Wang2018h,Chang2016,Weigel2018}. In the latter one, the engineering and design of ferroelectric domain structures plays a crucial role, as domain grids can be used to obtain phase matching between the interacting beams in nonlinear optical frequency conversion processes by employing the quasi-phase matching (QPM) technique. Phase-matching is required for any efficient nonlinear optical interaction. These processes are a cornerstone in many applications, such as for frequency conversion between different optical bands, pulse compression, LIDAR, spectral filtering, or in single photon sources in quantum optics \cite{Allgaier2017,Sharapova2017,Imeshev2001,Garmire2013,Fernandez-Gonzalvo2013,Maring2018}. However, the key to QPM is the fabrication of high quality ferroelectric domain grids over a millimeter to centimeter scale, which requires a thorough understanding of the underlying domain fabrication process, which is aided by visualization and analysis tools such as SH microscopy.

One of the most common ferroelectric materials for nonlinear, integrated optics has been lithium niobate (LN), due to its chemical and physical stability, wide availability, large transparency window and large nonlinear and electro-optic coefficients \cite{Weis1985}. Optical waveguide structures in bulk LN are fabricated by diffusion methods, which achieve low transmission losses ($<0.1$~dB/cm) \cite{Bazzan2015,Luo2015}, but offer only weak confinement, leading to large mode sizes, large bending radii and ultimately limited efficiency in frequency conversion devices. The recent commercial availability of LN on insulator (LNOI), however, addresses this issue by enabling sub-wavelength confinement due to large refractive index contrasts while preserving the special properties of LN\cite{Boes2018,Rao2018}. A typical cross-section of a LNOI wafer is displayed in Fig.~1. LNOI consists of a thin film of LN, which is bonded to an SiO$_2$ layer (buried oxide, BOX) deposited on a handle, e.g. LN or silicon. The LN thin film typically has a thickness between 300 nm and 1000 nm, while the BOX has thicknesses of 3~$\mu$m or less \cite{Boes2018,Rao2018}. The high vertical refractive index contrast allows the fabrication of low loss ($<0.1$~dB/cm)\cite{Desiatov2019}, single mode and sub-wavelength confinement waveguides with various methods, e.g. etching\cite{Desiatov2019,Liang2017}, ridge loading \cite{Rao2017}, diffusion \cite{Cai2015a,Cai2016a} or in a hybrid approach, where the thin film is die-bonded to prefabricated silicon photonics wafers \cite{Weigel2016,Weigel2018e,Chang2016}.

\begin{figure}
	\includegraphics[width=0.9\linewidth]{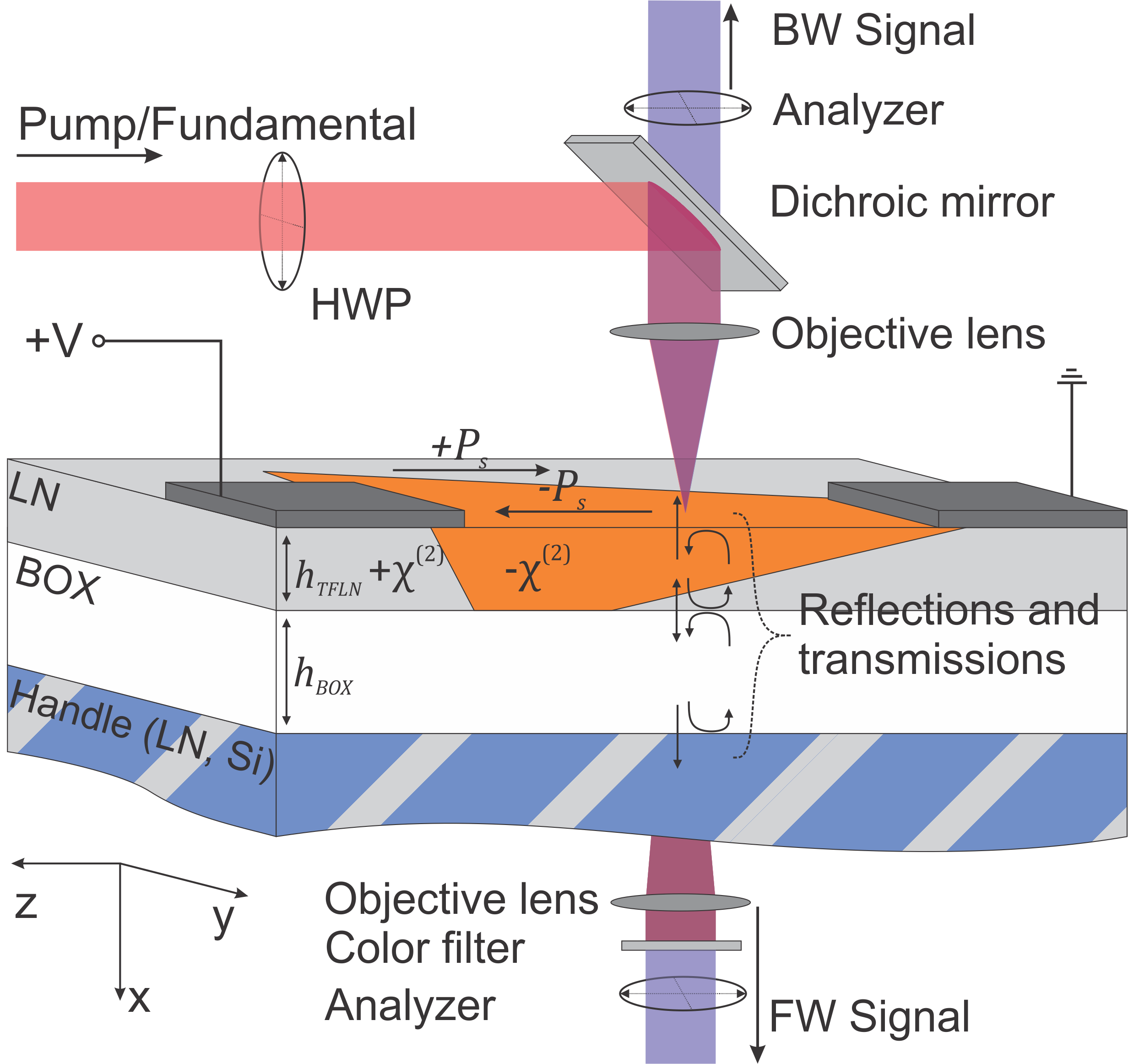}% Here is how to import EPS art
	\caption{\label{fig:asdasd} Typical geometry for the SH microscopy investigation of poled x-cut LNOI.}
\end{figure}

Recently, many high-efficiency frequency conversion devices have been demonstrated in periodically poled x-cut thin films \cite{Wang2018h,Chang2016}. This geometry allows access to LN's largest nonlinear tensor element $d_{33}$ with 
single mode waveguides, which guide the transverse-electric polarization, in which the electric field vector of the guided mode is oriented parallel to the waveguide substrate interface, i.e. the z-axis shown in Fig.~(1), which is identical to the optical axis of the LN crystal. In principal other crystal orientation of LN can be used, such as z-cut. However, the fabrication of QPM structures in z-geometry requires buried electrodes, which can complicate low loss waveguide fabrication due to the electrode material buried below the film. Furthermore, it is still under investigation if ferroelectric domains in z-cut thin film LN with less than 1~$\mu$m thickness are long-term stable\cite{Volk2017} or not \cite{Shao2016}. In contrast, x-cut films have been successfully poled with lithographically-structured, deposited electrodes on the top of the film, such as shown in Fig.~1 \cite{Mackwitz2016,Zhao2019,Wang2018h,Chang2016}. Poling electrodes like this can either be removed after poling or fabricated with sufficient distance to any guided optical modes for low propagation loss \cite{Zhao2019,Chang2016}.

Efficient frequency converters require highly uniform domain structures with respect to the respective design parameters. Often, a domain grid with one specific poling period and a  50\% duty cycle is desired with vertical DWs with respect to the waveguide mode propagation direction. The poling process, however, is characterized by a strong anisotropy: inverted domains nucleate under the application of a poling pulse on the positive electrode growing rapidly towards the negative electrode and spread (at a slower speed) in the lateral directions. In thin films poled with top electrodes, this can result in domain structures which are broad or overpoled close to the positive electrode, while not being inverted in depth along the complete length between the poling electrodes. A schematic diagram of such a structure is shown in Fig.~1. An important question for any thin-film domain imaging tool is whether vertical domains walls, completely or incompletely inverted domain structures can be distinguished. 

The SH contrast mechanism for bulk domain structures is reasonably well understood and many observations can be explained by interference and phase-(mis)-matching arguments \cite{Kaneshiro2010,Huang2017}.  However, the SH signal in thin films is influenced by various parameters, such as co- and counter-propagating phase matching, resonant enhancements, (resonant) reflections or absorption. 

Recently, we suggested that the signal detected in back-scattering direction for thin films of LN on a silicon handle is mainly due to reflected SH light, which is phase-matched in co-propagation \cite{Zhao2019}. In principle, this suggests that the second harmonic signal can unambiguously allow us to distinguish between completely and incompletely inverted domains and, therefore, can potentially be used to measure the depth of an inverted domain, with a depth resolution that is beyond the diffraction limit. Therefore, in this work we have investigated the SH imaging process in thin films of LN with simulations to quantify and understand the effects of the various influence parameters.

This paper is structured as follows: Section II explains the experimental and theoretical considerations of SH microscopy in thin films including phase-matching and explains the simulation setup. The main simulation results are presented in Sec. III, which is separated into two parts. In Part~A, we explore the geometrical and experimental influence factors on the SH signal, while in Part~B, we focus on experimental relevant theoretical analysis of domain imaging, such as the determination of domain depth or simulated line scans. In Sec. IV, the simulation results are compared with experimental results. Concluding remarks are given in Sec. V.

\section{\label{sec:Model}Methodology and physical principles}

\begin{figure}
	\includegraphics[width=0.8\linewidth]{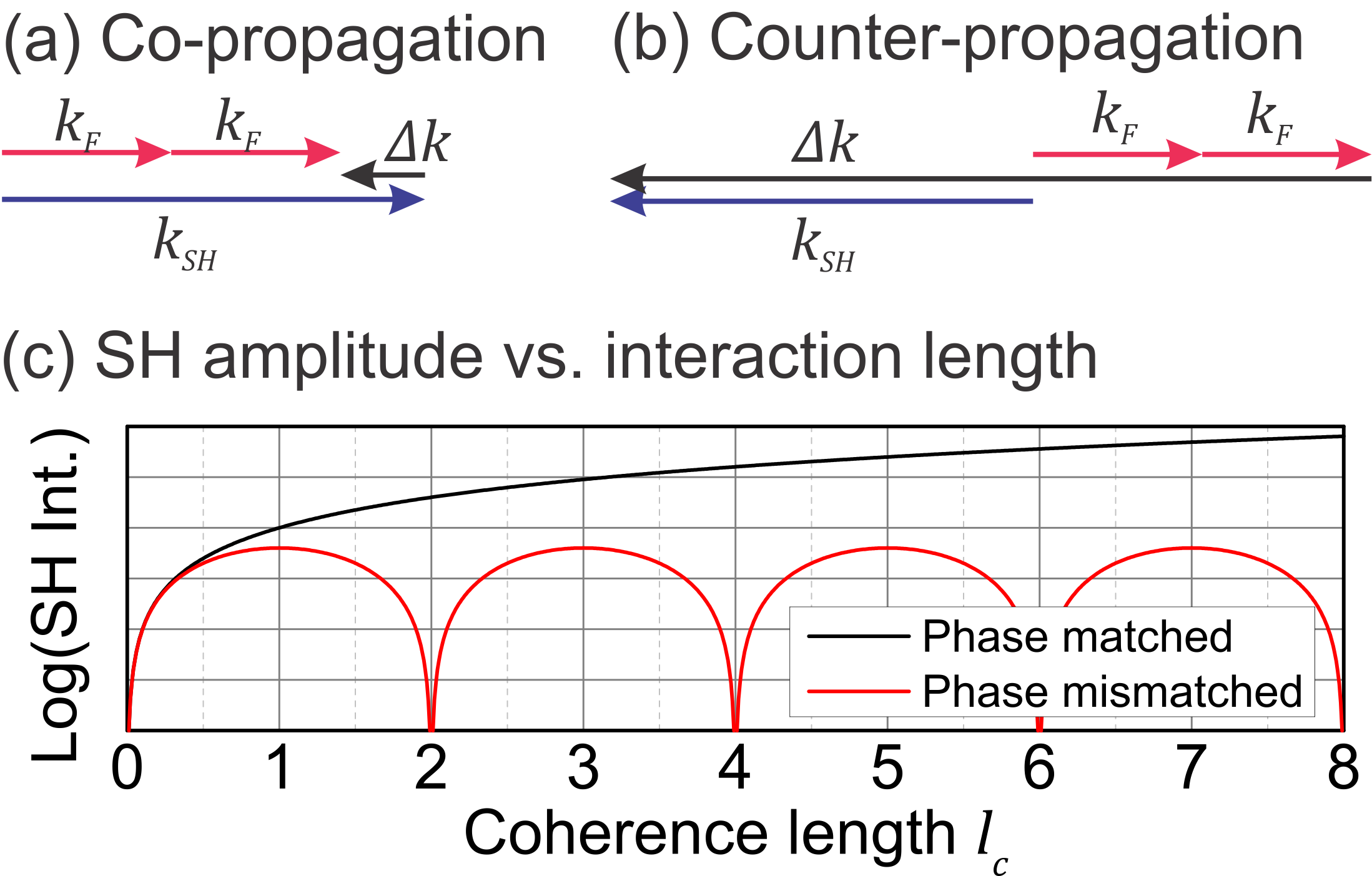}% Here is how to import EPS art
	\caption{\label{fig:asd} Vector diagrams for the SH generation process for (a) co- and (b) counter-propagating beams. (c) Generated second harmonic power as a function of interaction length for a phase matched versus a phase-mis-matched process.}
\end{figure}

A typical geometry for second harmonic microscopy is depicted in Fig.~1. Light from a high peak-power laser, typically a mode-locked laser providing femtosecond-pulses, is focused to a diffraction-limited spot, by to high numerical aperture (NA) objective enabling high spatial resolution. The SH light is generated at one-half the pump wavelength and is either collected in backward (BW) direction, i.e. by the same objective, or in forward (FW) direction by another objective, providing the sample is transparent at the respective wavelength \cite{Berth2007,Huang2017}. The pump light is then blocked in the detection path by filtering and/or wavelength selective attenuation. The (usually weak) SH light is detected by means of single photon counting. Usually, the polarization of the pump and SH light is set by polarizers, because the SH generation is highly polarization selective. As this technique requires point excitation and detection, scanning of the sample is necessary to generate an image. In many setups, this is either performed by scanning the beam, e.g. by galvo mirror systems, or by means of a fixed focus and scanning of the sample, which is mounted on a piezo stage.

The experimental setup employed in our experiment uses a femtosecond mode-locked Ti:Sapphire laser as the pump source (KMLabs Griffin) with a center wavelength around 800~nm and pulse length shorter than 100~fs. As different samples generated SH light with different efficiencies, pump powers between 5 to 150 mW are used to generate sufficient signals for integration times between 5~ms and 50~ms. For focusing, we employ two different microscope objective lenses (50x Mitotuyo Plan Apo, infinity corrected, NA~$=0.55$; 100x Zeiss Epiplan-NEOFLUAR NA~$=0.9$). In the samples we studied the SH light can only be detected in BW direction, collected by the same objective lens. Raster scanning is provided by scanning the sample with respect to a fixed focal point, using via a piezo-driven nano-positioning unit (PI Nanocube) to which the sample is mounted. The generated SH light around~400 nm is filtered by a dichroic mirror and appropriate color-glass filters (Schott BG39) and detected via a silicon single photon avalanche diode (MPD, PDM series). More details about the setup can be found elsewhere \cite{Zhao2019}.

\begin{table}
	\begin{ruledtabular}
		\caption{\label{porto} Refractive indices of the materials used for the calculations for the pump wavelength of 800~nm and its SH \cite{Apnes1983,Zelmon2008,Malitson1965,Green2008}.}
		\begin{tabular}{cccc}
			Wavelength  & LN(e)  &  SiO$_2$ & Si \\
			800~nm (Fund) & 2.1755 & 1.4533 & 3.6941+0.0065435$i$ \\
			400~nm (SH)& 2.3321 & 1.4701 & 5.5674+0.38612$i$ \\
		\end{tabular}
	\end{ruledtabular}
\end{table}

SH generation is only efficient if the interacting beams, i.e. the fundamental/pump (Fund) and SH beam, are matched in phase. Otherwise, the generated power measured along the axis of optical propagation will oscillate with a period determined by the coherent buildup length $l_c$, also called to the (nonlinear) coherence length. The coherent built-up length $l_c$ is the maximum length to which a power built up is observed, before phase mismatch between the pump and the generated wave results in amplitude decrease rather than growth. Figure 2 shows the $k$-vector diagram for a collinear second harmonic process for (a) co-propagation, i.e. the generated SH $k_{\textnormal{SH}}$ and the two photons of the pump light $2k_{\textnormal{Fund}}$ are propagating in the same direction, and (b) counter-propagation, where the SH light is generated in the opposing direction of the pump light travel. The phase mismatch $\Delta k$ can be calculated by

\begin{equation}
2k_{\textnormal{Fund}} - \Delta k = \pm k_{\textnormal{SH}}.
\end{equation}

\noindent Here, $+k_{\textnormal{SH}}$ and  $-k_{\textnormal{SH}}$ denote co- and counter-propagation, respectively. In general the phase mismatch between the pump and fundamental beam is determined by the dispersion of the refractive index $n$ through $k=2 \pi n /\lambda$, where $\lambda$ is the respective wavelength. With $\Delta k = \pi/l_c$, \cite{Boyd2003} we obtain

\begin{equation}
	 l_c = \frac{\lambda_{\textnormal{Fund}}}{4}\frac{1}{n_{\textnormal{SH}} \pm n_{\textnormal{Fund}}}
\end{equation}

\noindent for the coherence length $l_c$. In Eq.~(2) the positive sign denotes counter-propagation, while the negative sign denotes co-propagation. Assuming a SH process with an 800~nm pump wavelength, we calculate a coherence length for counter-propagation of $l_{C,counter} \approx 44$~nm and for co-propagation of $l_{C,co} \approx 1.28$~$\mu$m based on the extraordinary indices of LN given in Tab. 1. Typical LNOI film thicknesses are less than 1000~nm. The second harmonic power scales quadratically with interaction length \cite{Boyd2003}. This means that SH light generated in co-propagation will be up to two or three orders of magnitude stronger than in the counter-propagating case due to the 10 to 20 times longer interaction length. 

In microscopy of bulk LN samples, the co-propagating signal may only be detected, if a second microscopy objective in FW direction is used \cite{Berth2007,Huang2017}. However, the thin film structures includes several reflecting interfaces within the optical depth resolution. For plane wave incidence the reflectivity of any interface can be calculated from Fresnel's equations as

\begin{equation}
R = \left|\frac{n_1 - n_2}{n_1 + n_2} \right|^2.
\end{equation}

Taking the refractive indices of the involved materials into account, a reflectivity of up to $R_{\textnormal{Si/BOX}} = 0.34$ can be calculated for 400 nm light at the silicon/BOX interface. Even when a LN handle is used, a reflectivity of $R_{\textnormal{LN/BOX}} = 0.05$ will be observed. For typical film thicknesses of a few hundred nanometers, the SH light generated in co-propagation is two orders of magnitude more intense than the counter-propagating signal. Therefore, even after reflection and transmission losses, the SH signal in BW detection will primarily consist of reflected co-propagating SH light, rather than the directly-collected counter-propagating light. To avoid confusion we will refer to the detection geometry by either FW or BW detection, while the phase matching process will always be referred to as either co-propagation or counter-propagation.

The signal strength will further be influenced by resonance effects from the layer structure, either in the film or the BOX. The resonance thickness $d_{\textnormal{res}}$ for a film for a given vacuum wavelength $\lambda_0$ and refractive index $n$ can be calculated by

\begin{equation}
 d_{\textnormal{res}} \cos(\Theta) = m \frac{\lambda_0}{2 n}.
\end{equation}

In a highly focused beam, the center parts of a beam will propagate orthogonally to the surface, while the outer parts will refract into the surface. Therefore, an ensemble of propagation directions is present in any focused beam. For higher NAs this may lead to broadened and smeared out resonances, as different parts of the beam are resonant at other conditions.

\begin{figure}
	\includegraphics[width=0.9\linewidth]{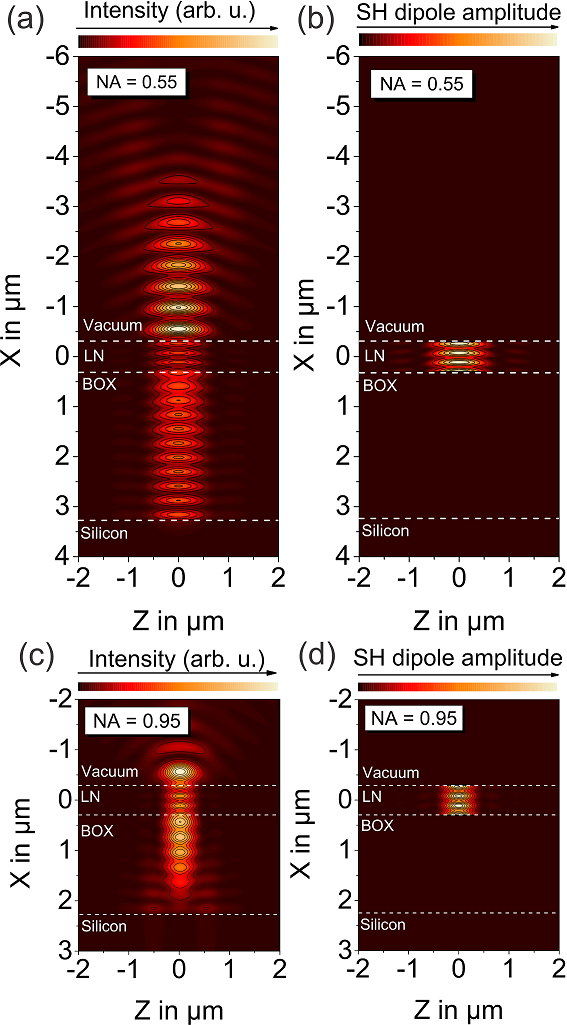}% Here is how to import EPS art
	\caption{\label{fig:caasdasdan} Calculated focus intensity distributions for an objective lens of (a) NA~$=0.55$ or (c) NA~$=0.95$ on a LNOI structure with $h_{\textnormal{LN}}=0.5$~$\mu$m and (a) $h_{\textnormal{BOX}}=3$~$\mu$m or (c) $h_{\textnormal{BOX}}=2$~$\mu$m. The slices are taken through the center of the focus $Y=0$. For both cases a strong influence of resonances and reflections on the focus field distribution can be seen. Panels (b) and (d) show the respectively induced SH dipole amplitudes, which only are visible in the LN film.}
\end{figure} 

The explanations so far have provided mainly a qualitative insight on the main physical principles, which influence and change the SH signal response in LNOI structures. To provide a more detailed and quantitative analysis on these influence factors, we have performed numerical calculations. For this, we employ a model and code based on work by Sandkuijl et al. \cite{Sandkuijl2013,Sandkuijl2013a}. The code is provided for Matlab and available under GNU General Public License \footnote{D. Sandkuijl, Computational code for second and third harmonic generation in layered media with high numerical aperture focusing, http://hdl.handle.net/1807/32992}. The SH response is calculated using the following steps: First, the focal spot distribution is calculated in a full three-dimensional calculation for a given NA of the focusing lens at the pump wavelength. This, for example, enables us to accurately describe axial polarization components in the focus region, which may be large for tight-focusing conditions \cite{Saito2008}. The simulation setup allows us to arbitrarily choose an input field at the back of the lens aperture, e.g. higher order modes or azimuthal or radial polarization modes are possible. However, any calculations in this work were performed with a linear polarized, fundamental Hermite-Gauss mode, $HG_{00}$. The focal field distribution will be influenced by (multiple) reflections or absorptions in thin film structures as shown in Fig.~1. We account for these effects using a transfer matrix formalism in the focal spot calculations. We account for absorptions using a complex refractive index. In the current form, the code can only include reflections from layers perpendicular to the optical axis of the beam, and extending infinitely in the horizontal plane. This is a good assumption for microscopy on films, where the film extension is significantly larger than the lateral extension of the focus. Two examples for focal spot calculations in the LNOI geometry are shown in Fig.~3(a) and (c). In both examples a strong influence of reflection and the formations of standing wave-like patterns can be seen. Only some light penetrates into the silicon layer due to high reflectivity and significant absorption. In the second step, the locally induced SH dipole strength is calculated based on the intensity distribution at the three polarizations and the nonlinear tensor $\chi^{(2)}$ distribution. An example for this is shown in Fig.~3(b) and (d), where a SH dipole amplitude is only induced in the area representing the LN film. The distribution can be chosen arbitrarily to represent any three-dimensional distribution independent of the interfaces. For example, this is used to analyze the signal from a thin film on a handle with or without a $\chi^{(2)}$ nonlinearity, or to analyze wedge-shaped domain structures, as indicated in Fig.~1. In the third step, the local SH dipoles radiate into the far field in FW or BW directions. At this step again, we account for reflections and transmissions using a transfer matrix formalism. In this step, phase matching is also accounted for, as the dipole radiation will be summed phase-correctly in the far field. Finally, the intensity distribution at the back of the collecting objective in FW or BW direction is calculated inversely to the first step. This last step allows us to compare differences or similarities in BW and FW signals. To obtain an intensity similar to an experimental signal, we calculate the integral of the normalized and squared field distribution at the back of the BW and FW apertures for the two main polarizations in our simulation. Details about the implementation can be found in the original works by Sandkuijl et. al. \cite{Sandkuijl2013,Sandkuijl2013a}.

For all of our calculations, the input mode is a linearly polarized, fundamental Hermite-Gauss mode $HG_{00}$, which is overfilling the back aperture of the objective. For any practical purposes, this is equivalent to an objective lens illuminated with plane waves and ensures the smallest, diffraction-limited focal spot in the calculation. The linear polarization was chosen to be parallel to the z-axis of LN as shown in Fig. 1. This allows us to address the largest nonlinear tensor element $d_{33}$. Similarly, only the z-polarized component was considered on the detection side, which models the experimental geometry, where the incident polarization and the detection polarizer is chosen parallel to the z-axis of LN in order to ensure only contributions by the $d_{33}$-element. If not otherwise stated, e.g. in wavelength sweeps, the pump wavelength is set at 800~nm. The calculation is performed on a grid of usually 20 nm in lateral direction and 10~nm spacing in axial direction, which is better than $\lambda_{SH}/10$ in the LN film and better than $\lambda_{SH}/20$ in SiO$_2$. A better grid spacing was chosen, if sweeps required the variation of a parameter with higher resolution, e.g. the film thickness. This keeps the usage of computational resources and calculation time at a manageable level to perform sweeps of various parameters. The film center is placed at the origin of the coordinate system, which equals the center of focus in the calculations. The code currently does not account for birefringence, which can lead to substantial focus distortion and/or change in phase matching \cite{Jain2006a,Gusachenko2013}. However, in our calculation we are only detecting in one polarization, i.e. addressing the same refractive index. Further, we are primarily analyzing the signal from a thin film, where focus distortions due to birefringence practically play no significant role, which only start to become dominant when focusing tens or hundreds of microns into birefringent media \cite{Jain2006a}. The absorption in silicon can be accounted for with a complex-valued refractive index. However, to compare some results of the FW and BW scattered, light we have set the imaginary part $\kappa_{Si} = 0$ in the simulations. Because the imaginary part has only a small influence on the absolute value of Eq.~(3), this induces only a very small error, on the order of 2\% in the calculated signal. In our simulation, only the $d_{33}$ element of lithium niobate parallel to the z-axis is chosen to be non-zero and the pump light is linear polarized parallel to this orientation. This is a reasonable assumption, because the value of $d_{33}$ is more than 5 times larger than the next largest tensor element ($d_{33}=$~34~pm/V; $d_{31}=$~6~pm/V; $d_{22}=$~3~pm/V)\cite{Nikogosjan2005}. Further, in the experiment and the calculation, we only detect light polarized parallel to the z-axis. In this configuration, other elements and light polarization can only be addressed by field polarization components present due to the strong focusing. For a NA~$=0.9$ the axial field element can account for up to 20\% of the total intensity  \cite{Saito2008}. But as the SH power scales quadratically with the tensor element and pump power the signal generated through interactions with other tensor elements is low even for NA~$=0.95$ (less than 1\%). In contrast, in the investigation of z-cut LN significant SH signal may be generated by the axial field component. This can make interpretation of SH results and signatures challenging in z-cut geometry, and can require the use of lower NA, i.e. sacrificing spatial resolution, in exchange for a clear nonlinear signature. However, since in x-cut LN the largest element $d_{33}$ can be addressed directly, the signal is almost exclusively generated through the interaction with the $d_{33}$ element. This allows us to perform the experiments using the highest numerical aperture objective, and achieve the highest in spatial resolution. It should be noted, the assumption only to chose one nonlinear tensor element to be non-zero is not be valid anymore, if different types of input beams or polarizations are chosen, e.g. a radial polarized beam is used intentionally in order to achieve a very large axial polarization component, or if microscopy is performed on other orientations of LN film (e.g. z-cut).

\section{\label{sec:results}Simulation results}

\subsection{\label{sec:influence}Understanding the SH signal from LNOI}

\begin{figure}
	\includegraphics[width=0.95\linewidth]{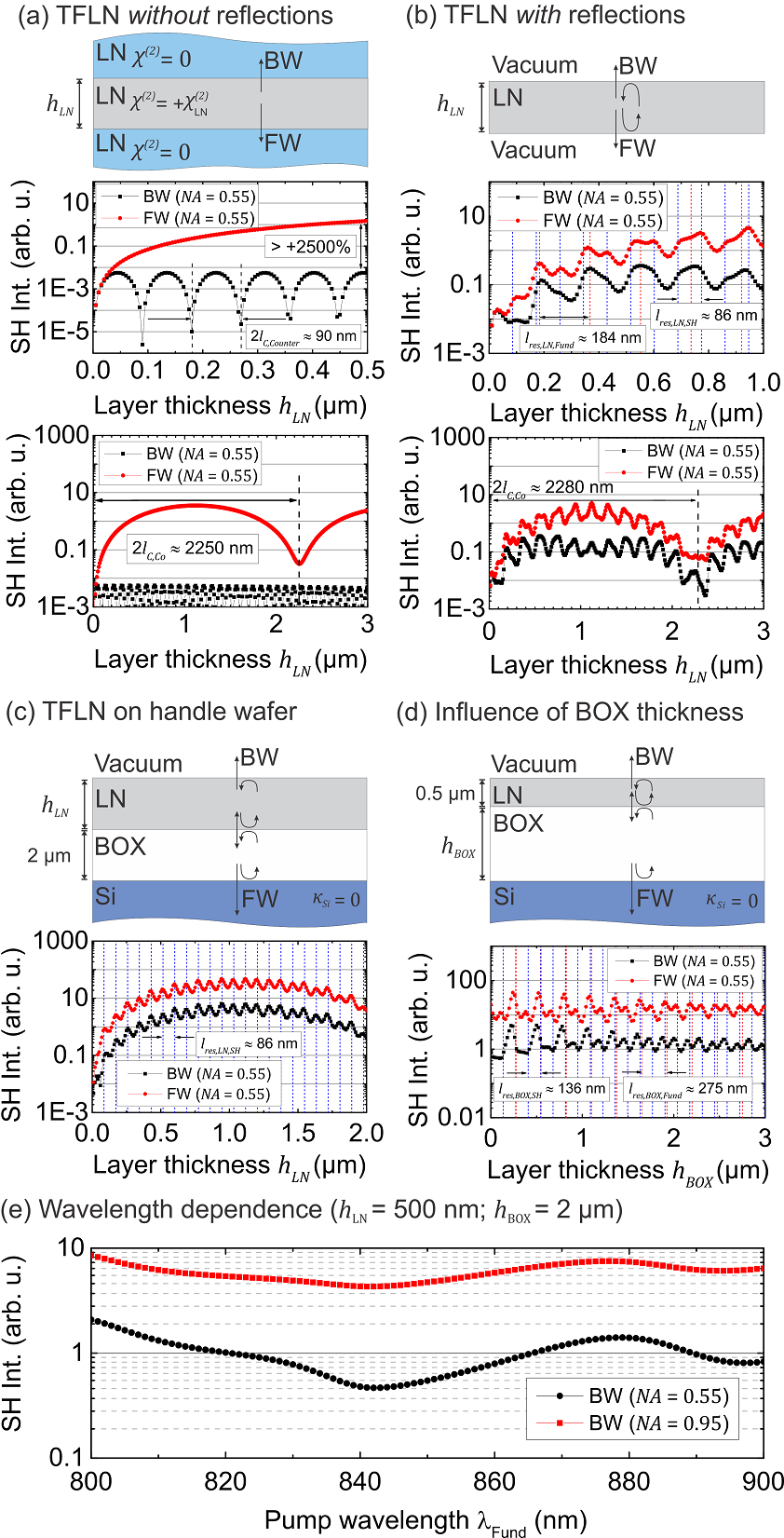}% Here is how to import EPS art
	\caption{\label{fig:an} Simulated influence of different parameters on the SH microscopy investigation of LNOI. (a) Thickness dependent SH signal from a LN layer with no reflections or interfaces and a background index equivalent to that of LN. (b) Thickness dependent response from an LN film suspended in vacuum ($n=1$). (c) Thickness dependent response from an LN film on an LNOI structure. (d) Influence of the BOX thickness on the nonlinear signal from an LNOI structure. (e) Wavelength dependent SH response from an LNOI structure of the given parameters.}
\end{figure} 

The SH signal generated in thin films is influenced by various parameters, such resonances, reflections, phase matching in co- or counter-propagation, in poled or single domain structures. The influence of these parameters, in turn, will depend on the wavelength, NA and the thicknesses of the layers. The most basic simulation of SH radiation from a thin film comprises just a nonlinear film with given refractive indices, but \emph{no} reflections. This provides a baseline for the subsequent calculations which include reflections. For this, we have simulated the SH response from a nonlinear film with the refractive index of LN, which is suspended in a background of the same refractive index, but no nonlinearity in the background. The geometry is shown in the top panel of Fig.~3(a). The focus is placed at the center of the nonlinear film. In the simulation, the thickness of the nonlinear film was gradually increased and the SH intensity was calculated for FW and BW scattering for each thickness. The result is present in the middle and bottom panel of Fig.~3(a). The middle panel shows the result for the layer thickness for $h_{\textnormal{LN}} = 0$ to $500$~nm in 5~nm steps for BW and FW scattering and an NA of 0.55, while the bottom panel presents a zoomed out range of only FW scattered light for the same NA calculated in 10~nm steps to a thickness of up to 3~$\mu$m. As expected, the BW signal oscillates in a sine-squared function with an approximate period of 90~nm, which fits well with the calculated coherence length $2l_{C,Counter}= 88$~nm for counter-propagation. On length scales shorter than $500$~nm, the behavior of FW and BW radiation resembles the case of  phase-matched versus not-phase-matched interactions, as shown in Fig.~2(c). In contrast to the BW signal, the FW light shows a coherence length of more than 1~$\mu$m. Because the SH intensity scales quadratically with interaction length, a more than two orders-of-magnitude higher intensity is observed for a 500~nm interaction length. On longer scales, the FW signal also shows a sine-squared oscillation. From the simulation we inferred an approximate coherence length of $2l_{C,Counter}= 2250$~nm, which is slightly shorter than previously predicted value of $2l_{C,co} \approx 2550$~nm. It should be noted that the observed coherence length is shorter due to the use of focused light. Due to focusing, components in the pump beam propagate with orientations not-quite orthogonal with respect to the surface. This effectively increases the interaction length especially for the outermost parts of the focused beam. This leads to a decreased observed coherence length for larger NAs and also a decrease in maximum contrast of the sine-squared behavior. For the counter-propagating process, this effect is small, as the film thickness is usually larger than the coherence length.

The absence of reflections means that the realistic experiments will diverge considerably from the above described. As when we now introduce reflections into the system, the observed response changes  drastically. Figure~4(b) shows the SH response from an LN film suspended in vacuum ($n=1$ for fundamental and SH wavelengths outside of the film). Similar to Fig.~4(a), the thickness of the film is varied and the intensity in FW and BW direction is calculated. Now, the signals in FW and BW directions show a different behavior compared to Fig.~4(a). The envelope of the FW signal follows the general enveloping same shape as in Fig.~4(a) and shows a minimum at a thickness of 2280~nm indicating the same coherence length $l_{C,co}$ for co-propation. However, the BW signal demonstrates a much higher intensity, being around 0.1 to 0.2 of the intensity the FW signal and following a very similar trend and shape. The intensity and similar shape of the BW signal can be explained by observing that the light scattered in co-propagation is reflected at the inner interface. According to Eq.~(3), the LN/vaccuum interface has a reflectivity of around $R_{\textnormal{SH}} \approx 0.16$ at 400~nm, which fits well with the observed intensity ratio. Introducing reflections also leads to resonant enhancements of the SH and fundamental light, which are both visible, as oscillations occur at $d_{\textnormal{res,LN,SH}}=86$~nm and $d_{\textnormal{res,LN,Fund}}=184$~nm length scales, respectively. The resonant enhancements are moderate, as the reflectivity is only $R_{\textnormal{SH}} \approx 0.16$. Therefore, the intensity is generally of the same order-of-magnitude to the non-reflection case in Fig.~4(a). The low reflectivity is also one reason for the relatively broad resonance peaks among the use of a focused beam, which leads to an ensemble of propagation directions being present in the beam, which may be resonant or not resonant at different angles.

In a typical LNOI structure, the LN film sits directly above a BOX layer, with a handle layer present below the BOX, as shown in Figs.~4(c) and (d). In Fig. 4(c), the thickness of the LN film $h_{\textnormal{LN}}$ for a fixed BOX thickness ($h_{\textnormal{BOX}}=2$~$\mu$m) is varied. For this simulation, the Si layer is simulated without absorption to compare the FW and BW signals. A result similar to Fig.~4(b) is observed. Due to the high reflectivity of the Si/BOX interface, and a resonantly enhanced reflection from the BOX layer as shown in Fig.~4(d) the intensity of both signals is even higher compared to Fig.~4(b), due to increased reflection and stronger resonant enhancements. In the signal an oscillation with a period of 86~nm can be seen, which fits well with resonance thickness for the SH light in LN. Here, in particular, the resonance of the SH light has a strong influence due to its higher reflectivity at the Si surface, because of the larger refractive index of Si at the SH wavelength. If the BOX thickness is varied for a given LN film thickness, as shown in Fig.~4(d), oscillations are also visible. Here, the SH signal is oscillating due to resonantly enhanced reflection from the BOX layer of pump and/or SH light. For this simulation a NA of 0.55 was chosen. A numerical aperture of 0.55 provides a depth resolution of $\approx 4.5$~$\mu$m. If a larger NA is chosen, a signal decrease due to increased depth resolution can be expected when a BOX thickness larger than the depth resolution is chosen, e.g. compare the focus sizes in Fig.~3 with Fig.~4.

As we have shown, the signal detected from a reflecting film heavily depends on resonances in the BOX or LN layer. Therefore, a wavelength dependence is expected in the measured SH signal. To analyze this influence, we have swept the pump wavelength from 800~nm to 900~nm for a film of $h_{\textnormal{LN}}=500$~nm and $h_{\textnormal{BOX}}=2$~$\mu$m for two numerical apertures. The wavelength dependent refractive indices are calculated based on Refs. \cite{Apnes1983,Zelmon2008,Malitson1965,Green2008}. For a numerical aperture of NA~$=0.55$, the choice of the wavelength can make a factor of 5 to 10 difference in the SH signal, while the resonant effects for the larger NA are less pronounced. This is because in a highly focused beam, an ensemble of different propagation directions is present at the same time. Parts of the focused light is resonant, while some parts are not due to Equ.~(4). It should be noted that stronger focusing leads to an overall enhanced signal, due to increased power density in the focus.

\begin{figure}
	\includegraphics[width=1\linewidth]{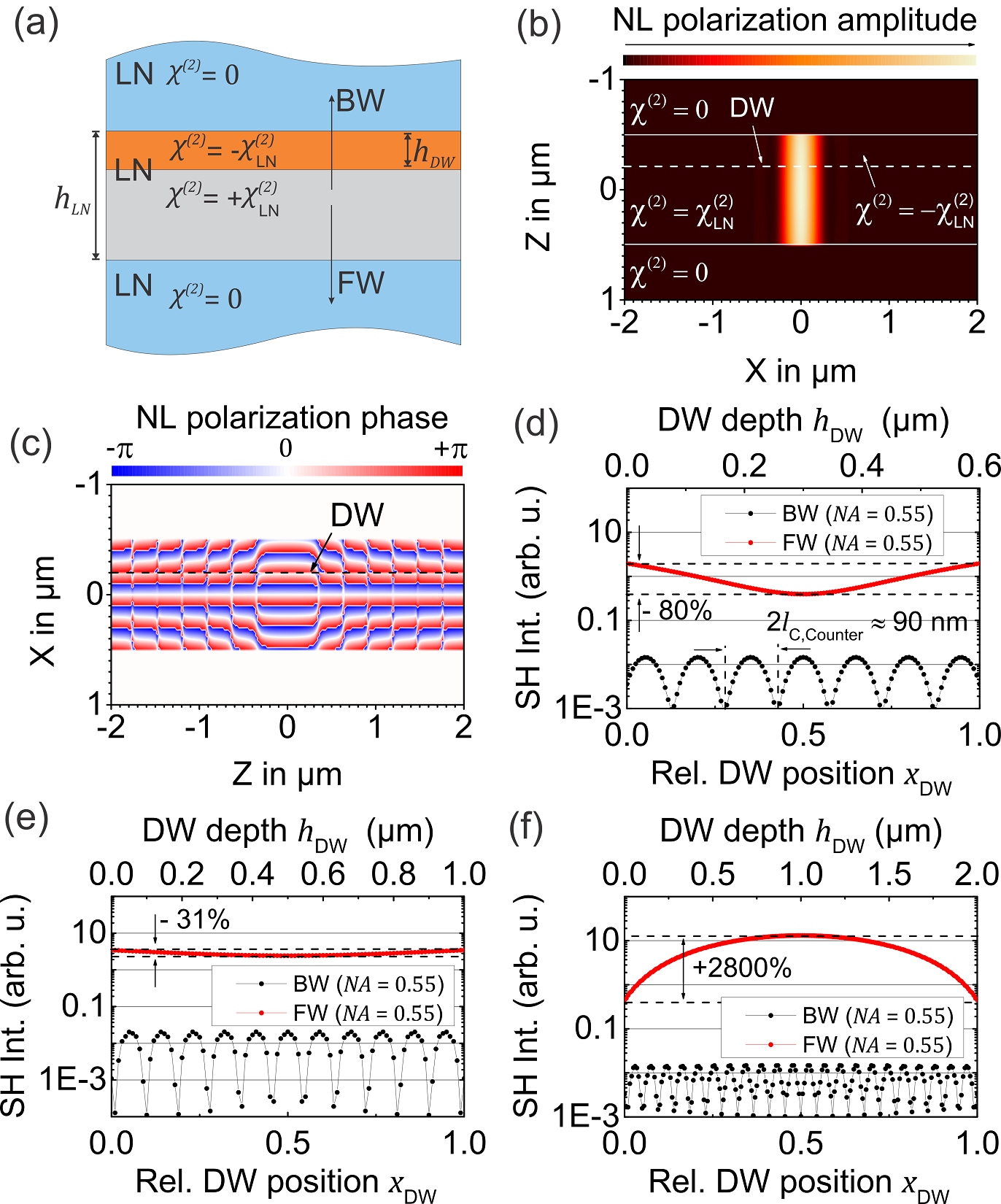}% Here is how to import EPS art
	\caption{\label{fig:4} Effects of inverted domains on the observed SH response from a film with no reflections. (a) Simulation geometry setup. The inverted domain is accounted for by a region of negative sign $\chi^{(2)}$. (b) and (c) Induced NL polarization amplitude in this structure. The amplitude is not affected by the DW transition, but in the phase an additional phase jump is visible. (d)-(f) FW and BW signal response for different film thicknesses, when the DW position is swept through the film.}
\end{figure}

The main motivation of SH microscopy on LNOI is to image and analyze ferroelectric domain structures. Domain structures can be simulated by choosing a specific $\chi^{(2)}$ distribution. If a region of LN is poled, the sign of the $\chi^{(2)}$ will be inverted, i.e. in a poled domain $\chi^{(2)}$ will be $\chi^{(2)} = - \chi^{(2)}_{LN}$. Due to the changed sign of the $\chi^{(2)}$-tensor, SH light generated in an inverted domain will be generated with a phase shift of $\pi$ compared to SH light generated in the non-poled part of the film. In this simple model, a DW is therefore a boundary between regions of $\chi^{(2)}$ with opposing signs. We do not assume an additional substructure of the DW region. For example, DWs have been observed to show additional symmetries \cite{Cherifi-Hertel2017} or effects of strain \cite{Rusing2018a,Dierolf2007}, which may contribute to additional or changed tensor elements. However, if such effects are desired to be modeled, such structures can be readily added into the model later. In Fig.~5, we have calculated the signal from films of different thickness, while a DW is gradually moved through the film. Similar to the example in Fig.~4(a), we have performed this calculation without reflections to first understand the effects of phase-matching on the FW and BW signals. Figure~5(b) shows the calculated NL polarization amplitude for a film of $h_{\textnormal{LN}}=1$~$\mu$m thickness and a DW depth $h_{\textnormal{LN}}=0.3$~$\mu$m. It should be noted that, in contrast to Fig.~3, no resonances are visible, as there are no reflecting interfaces present. Because the tensor element magnitude does not change at the domain transition, there is no difference visible on the local NL polarization amplitude. However, in the NL phase distribution in Fig.~5(c), the transition is visible as an additional $\pi$ phase shift at  $h_{\textnormal{LN}}=0.3$~$\mu$m. If this domain wall is gradually moved through the layer, the signals in FW and BW change. This has been calculated for a 600~nm [Fig.~5(d)], 1000~nm [Fig.~5(e)] and a 2000~nm [Fig.~5(f)] film. The top axis in each of these figures is scaled for the real DW depth, $h_{\textnormal{DW}}$, as shown in Fig.~5(a), while the bottom axis describes the relative domain position, defined as

\begin{equation}
x_{\textnormal{DW}} = \frac{h_{\textnormal{DW}}}{h_{\textnormal{LN}}}.
\end{equation}

For the 600 nm film the FW signal gradually decreases, until the DW depth is at one-half the film width, because the film thickness is less than the coherence length and the signal destructively interferes. For the 1000~nm film a similar behavior is observed; however, the contrast is decreased as the film thickness approaches the coherence length. In the 2000~nm case, the FW signal is heavily increased around $x_{\textnormal{DW}}=0.5$, because the coherence length is shorter than the film width. If the film is exactly half-poled, the SH signal generated in each half is constructively interfering (This is the QPM working principle). In contrast to the FW signal, the BW signal oscillates with a period of 90~nm in all cases. Here, a maximum is observed each time, when the poled part of the film is positively phase-matched with the unpoled film. For any film with a thickness greater than the coherence length in counter-propagation, the maximum intensity is independent of the film thickness. 

\begin{figure}
	\includegraphics[width=1\linewidth]{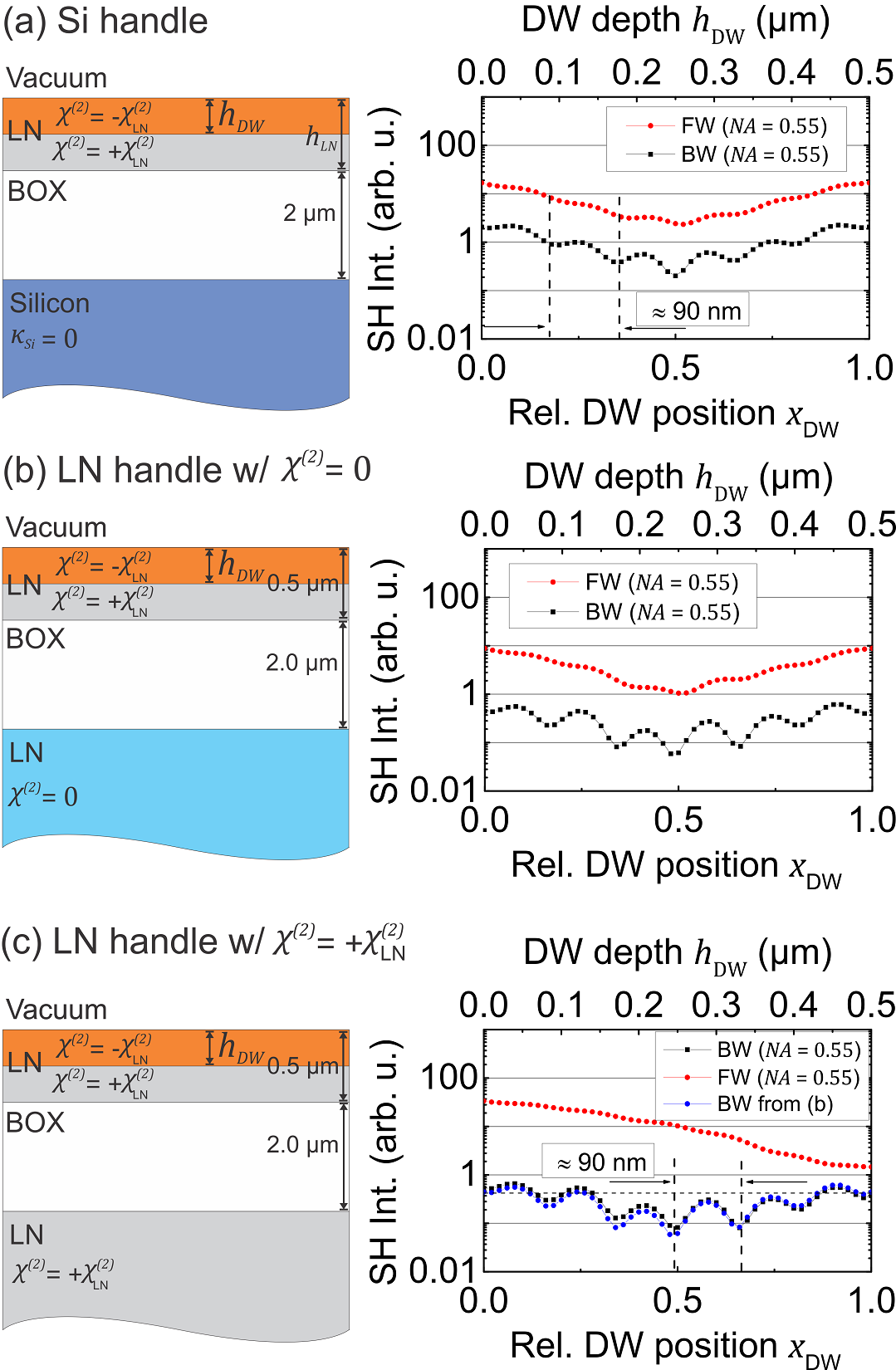}% Here is how to import EPS art
	\caption{\label{fig:5} Signatures of DW depth for LN films in an LNOI structure on (a) a Silicon handle, (b) a LN handle \emph{without} nonlinearity and (c) and LN handle \emph{with} nonlinearity and the same orientation as the unpoled film.}
\end{figure}

If now a poled film is placed on a BOX and Si handle structure as shown in Fig.~6(a), the same behavior for a gradually poled film is observed for the FW signal. The handle is simulated without absorption so that the FW signal can be seen. Due to the reflection the BW signal shows the same general characteristic where a minimum for a half poled film is observed. However, this signal is overlapped with oscillation with a period of 80-90~nm. It appears that these oscillations are due to counter-propagation phase matching in BW direction, and  contributions from resonance enhancements of the SH light, which has a similar oscillation period of 86~nm. Since the counter-propagating light is also reflected, a weak oscillation is also present in the FW signal. But the contrast is lower due to the stronger co-propagating signal. If the film is placed on a LN handle, which shows only reflection but not nonlinearity, as shown in Fig.~6(b), almost the same signal behavior is observed, except that the intensity of both the FW and BW signals are a factor of 5 to 10 weaker than observed for a silicon handle due to the lower reflectivity and weaker resonances. The relative height of the oscillations is increased due to weaker reflection of the co-propagating signal. If the nonlinearity of the handle is accounted for, the situation changes, especially for the FW signal. The BW signal is minimally influenced and shows a similar overall behavior compared with Fig.~6(b), but some changes, with differences up to a factor of 2, are visible due to interference with counter-propagation SH light generated in the handle. In an experimental context, this can be significant. In particular, the FW scattered light is greatly altered when the handle's nonlinearity is accounted for, because the co-propagating SH light interferes with SH light generated in the handle. This situation is similar to SH microscopy with reference samples \cite{Huang2017}, where a domain polarity contrast, rather than a DW sensitive contrast, is observed. Due to the stronger signal and clear behavior for inverted films, a microscope in FW direction may be advantageous for investigation of poled films on LN handles. However, the experimental study of poled films on Si handles at these wavelength must use the BW direction.

\subsection{\label{sec:poling_depth}Signal responses from poled LNOI}

\begin{figure*}
	\includegraphics[width=0.75\linewidth]{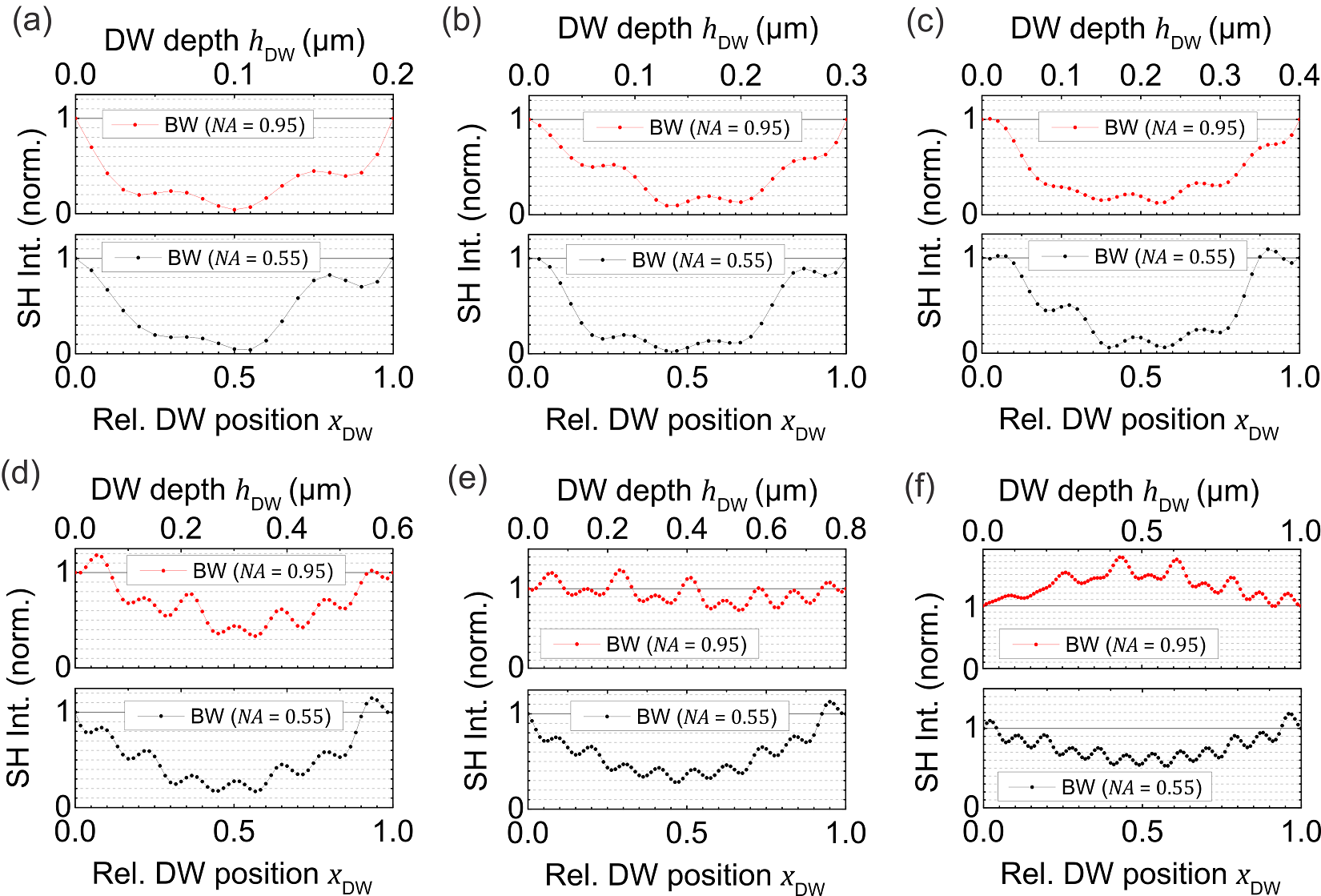}% Here is how to import EPS art
	\caption{\label{fig:7} SH signatures for a varying poling depth for LN films of different thicknesses on a silicon handle and a BOX thickness of $h_{\textnormal{BOX}}=2$~$\mu$m.}
\end{figure*}

\begin{figure*}
	\includegraphics[width=0.75\linewidth]{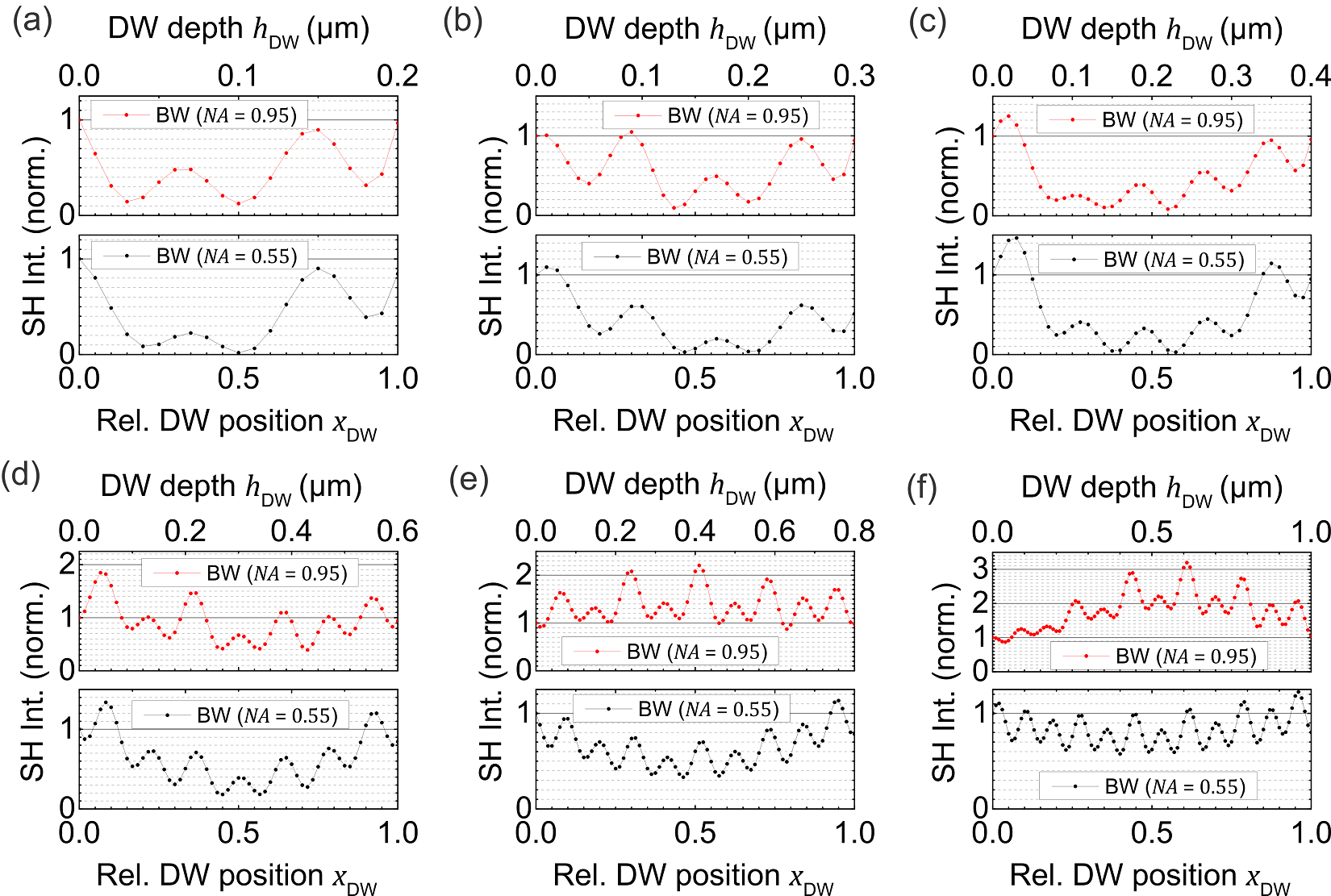}% Here is how to import EPS art
	\caption{\label{fig:8} SH signatures for a varying poling depth for LN films of different thicknesses on a LN handle and a BOX thickness of $h_{\textnormal{BOX}}=2$~$\mu$m.}
\end{figure*}

The previous discussion shows that the SH signal can be used to investigate the relative depth of poling in a film, and potentially allows us to unambiguously distinguish a film that is completely inverted in depth from a film that has only a shallow, near-surface domain structure. This is particularly important when x-cut films are poled with surface electrodes, because we seek to ensure that domains spread throughout the complete film thickness. Other standard methods, such as selective etching, are either destructive and require material ablation to procure a depth information, e.g. by focused ion beam, or are only sensitive to the sample surface layer, such as PFM, which is superior in lateral resolution compared to SH microscopy, but may only be sensitive to at-surface domain structures  \cite{Dierolf2007,Wittborn2002}.  

Therefore, we have investigated if the depth contrast in SH microscopy is observed for various film thickness. For this, we have performed calculations similar to Fig.~6(a) and (c) for film thicknesses between 200 and 1000 nm ($h_{\textnormal{BOX}}=2$~$\mu$m), as well as for two different NAs. The results are shown in Fig.~7 for the Si and Fig.~8 for the LN handle. It should be noted that, for the plots in this section, the calculated SH intensities have been normalized in each individual plot to the signal of an unpoled film ($x_{\textnormal{DW}}=0$). In an experimental context this is more relevant than the absolute intensities, because in a typical SH microscopy image of poled films, the raster-scanned area contains poled, as well as unpoled areas, which can serve as a reference point for the intensity. In our calculations, we observe, for thicknesses up to $h_{\textnormal{BOX}}=600$~nm, a gradual change in signal, with a minimum around $x_{\textnormal{DW}}=0.5$. For larger thicknesses, the contrast for a partly-inverted film decays when the film thickness becomes equal to the coherence length in co-propagation. This is very apparent for the highly focused beams, where the observed coherence length is shorter due to the strong focusing. For films of thicknesses greater than the coherence length a positive contrast is expected and observed, e.g. in Fig.~7(f). Oscillation with a period of approximately 90~nm, which were observed previously, e.g. Fig.~6, due to counter-propagation phase matching and resonance of SH light in the film, are now visible in all plots. Their magnitude is particular large for the LN handle (Fig.~8), where they show a significantly larger relative amplitude compared to the Si handle (Fig.~9). This is due to the lower reflection at the BOX/handle interface which leads to less light generated in a co-propagating process being detectable in the BW direction, and a relative stronger contribution from counter-propagation. The oscillations appear in particular strong for film of 800~nm and 1000~nm thickness, which are analyzed with a 0.95 objective. Here, no drop in intensity, but mostly oscillations with amplitudes of 2-3 times more intensity for a gradually poled film are predicted. These observation fit well with previous reports of SH microscopy on poled x-cut films visualized with large NAs, where instead of a homogeneous level highly oscillating signals have been observed \cite{Mackwitz2016}. Furthermore, for the LN handle, it is observed that unpoled ($x_{\textnormal{DW}}=0$) and completely inverted films ($x_{\textnormal{DW}}=1$) will not provide the same intensity. This is easily understood, as the SH light will constructively or destructively interfere with counter-propagating light generated in the handle. The effect is, in particular, visible for the films with smaller thicknesses, because the signal in co-propagation will be weaker compared to the counter-propagation intensity. For example, the SH signal for a poled film and unpoled films shows a difference of more than 50\% for the 300~nm film in Fig.~8(b) for the NA of 0.55 between the unpoled ($x_{\textnormal{DW}}=0$) and poled state ($x_{\textnormal{DW}}=1$). Therefore, for an LN handle, in principle, SH microscopy can also be sensitive for the domain polarity. From the previous section, it is known that the SH signal generated from thin films depends on the exact combination of wavelength, film and BOX thicknesses; slight variations on the order of less than 100~nm can change the SH microscopy signal by a factor of 5 to 10. While this will change the overall SH signal, it does not influence the fact that there is an intensity minimum in the BW signal for a partly-inverted film at approximately $x_{\textnormal{DW}}=0.5$. Slight changes in thickness of the BOX or pump wavelength, however influence the exact position of the 90~nm period oscillations changing the exact signal profile. 

\begin{figure*}
	\includegraphics[width=1\linewidth]{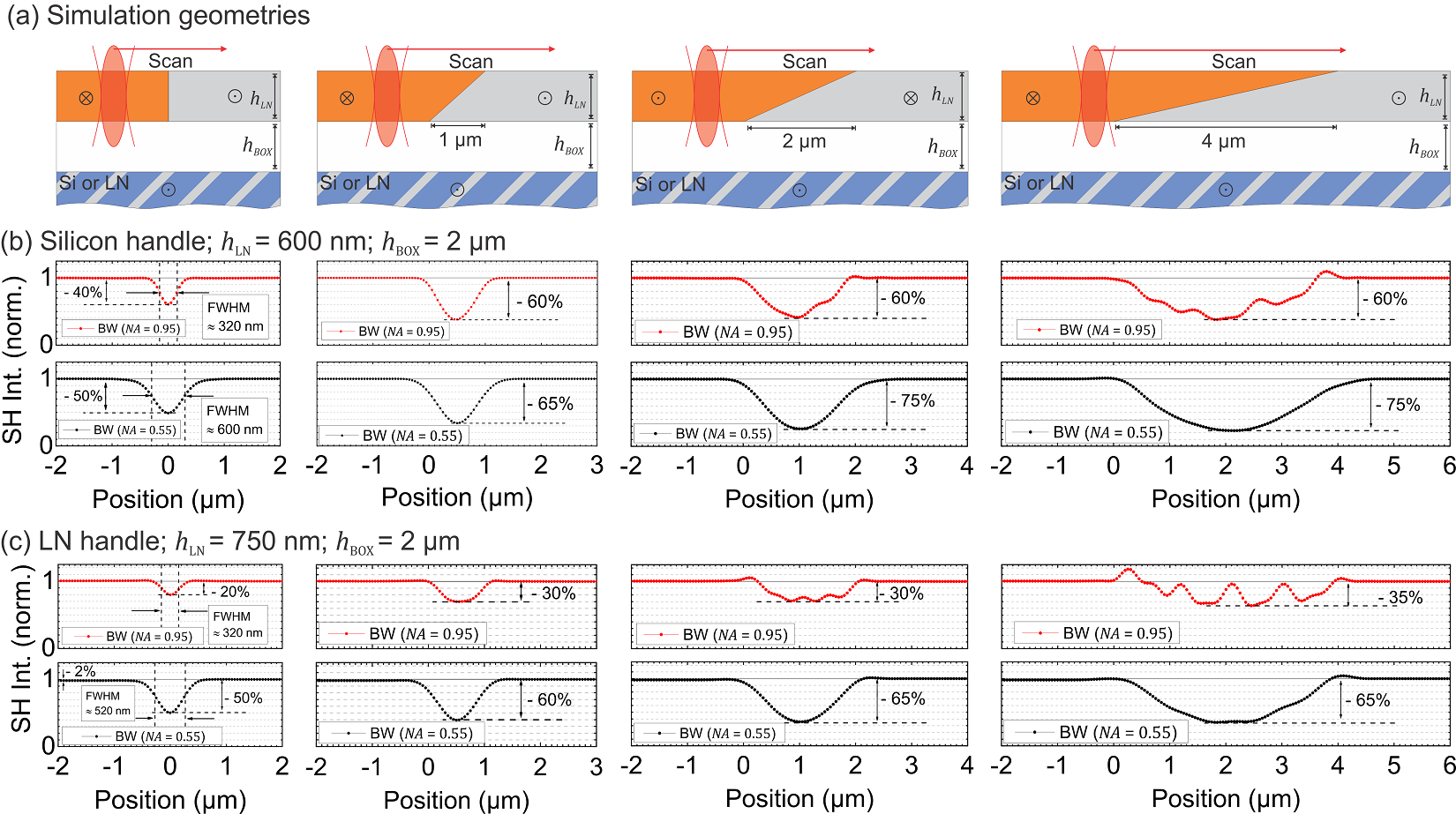}% Here is how to import EPS art
	\caption{\label{fig:9} (a) Calculation geometries for simulated line scans. (b) Results for a $h_{\textnormal{LN}}=600$~nm) film on $h_{\textnormal{BOX}}=2$~$\mu$m BOX and silicon handle. (c) Results for a $h_{\textnormal{LN}}=750$~nm) film on $h_{\textnormal{BOX}}=2$~$\mu$m BOX and lithium niobate handle.}
\end{figure*}

So far, we have simulated the effect of planar domain transitions extending infinitely in all directions. However, in a real sample, domain transitions may be vertical, rather than horizontal, with respect to the surface. This may be the desired geometry for a QPM structure with a waveguide oriented along the film. Potentially, domain structures may be angled with respect to the surface. In particular, angled or shallow domains are likely to be expected for x-cut LN, which is poled with top surface electrodes.  Poled domains in LN show a pronounced asymmetry in growth speed, i.e., the domain propagation in the z-crystal axis is two orders of magnitude faster than the spreading speed in the x-y directions\cite{Choi2012}, while the domains may form with a hexagonal cross-section in the xy-plane, as observed in various experiments \cite{Gui2009,Scrymgeour2005a}. 

Because microscopy is limited by lateral resolution, some of the previously calculated effects may be smeared out for typical samples, as a mixture of different depth is present in a lateral width of a focus. This may have a particular large effect on the the short length scale oscillations.  To analyze this, we have simulated lines scans for four different domain transition geometries as shown in Fig.~9(a). The first geometry is a vertical domain transition, while the domain transition in the second to fourth calculation is angled and stretched over 1, 2 and 4~$\mu$m, respectively. The z-axis and light polarization is orthogonal to the sketch plane. In the simulation, the respective structure is moved in 50~nm increments through the fixed focus and the intensities are calculated for each step. This process is very similar to the raster scanning performed in an experiment. The simulation is performed for two numerical apertures (NA~$=0.55$; NA~$=0.95$) for (a)  a silicon handle ($h_{\textnormal{BOX}}=2$~$\mu$m;  $h_{\textnormal{LN}}=600$~nm) and (b) a LN handle ($h_{\textnormal{BOX}}=2$~$\mu$m;  $h_{\textnormal{LN}}=750$~nm). For the vertical domain transition a drop in intensity can be seen when the focus is placed at the DW transition. In this configuration the signals generated in one domain destructively interfere with the signal generated in the opposing domain, leading to a decrease in intensity similar to the contrast on DWs in bulk samples \cite{Huang2017}. The width of this feature scales with the lateral resolution. The diffraction limited resolution for second harmonic microscopy \cite{Zipfel2003} given by

\begin{equation}
\Delta s \approx \frac{0.325}{\sqrt{2}}\frac{\lambda_{\textnormal{Fund}}}{\textnormal{NA}}.
\end{equation}

\noindent For NA~$ = 0.55$, we obtain $\Delta s = 340$~nm and for NA~$ = 0.95$ we obtain $\Delta s = 200$~nm. In the simulation, larger values are observed, as the focus width in the film structure is influenced by reflections, as seen in Fig.~3. For the angled domain structures, we see that the area of the decrease in intensity is enlarged accordingly, while the oscillations previously observed in Figs. 7 and 8 will appear to be particular strong for large NAs and only for more flat domains. The appearance of the oscillations provide an explanation why previous SH images of poled LNOI on a LN handle presented a marbled texture when imaged with large NA's\cite{Mackwitz2016}. It should be noted that, for the simulation in Fig.~9(c), we observe almost the same intensity (within a 2\% difference) for a completely poled or unpoled film on a LN handle. In this case, it is a coincidence that the intensities are the same for the chosen film and BOX thicknesses, and is not expected in general. As seen before (e.g. Fig. 8), the intensity levels for a inverted and not-inverted region on a LN handle are, in general, different.

\section{\label{sec:experiment}Experimental results}

\begin{figure*}
	\includegraphics[width=1\linewidth]{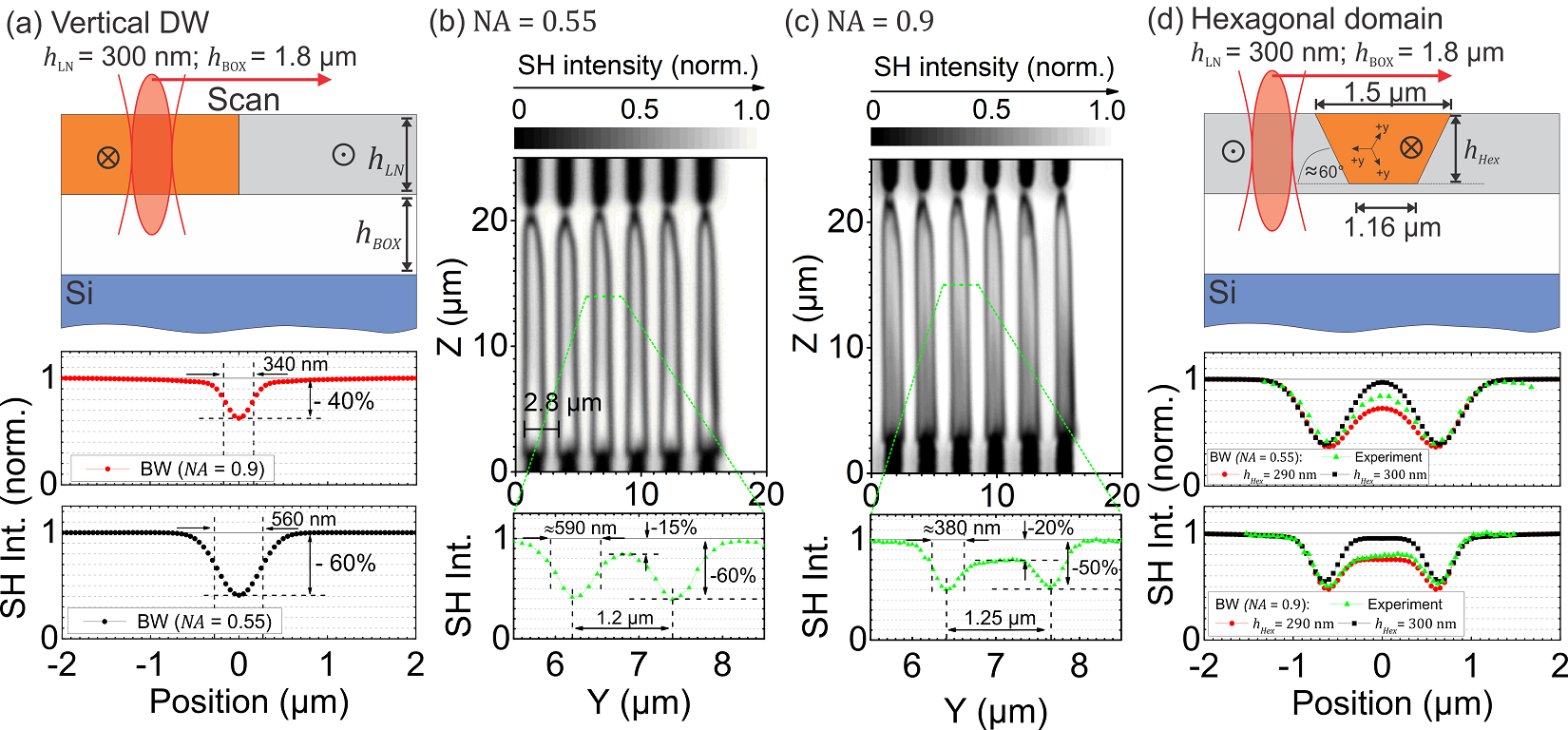}% Here is how to import EPS art
	\caption{\label{fig:10} Experimental investigation of a poled LN thin film with $h_{\textnormal{LN}}=300$~nm film on a BOX thickness of $h_{\textnormal{BOX}}=1.8$~$\mu$m and on silicon handle. (a) Simulated line scan for a vertical DW in this geometry. (b) and (c) Experimental results for two NAs on the same area of the sample. (d) Simulated line scans for a domain with a hexagonal cross-section.}
\end{figure*}

We have poled a sample of LN with lithographically-structured electrodes deposited on top of the film. The sample is a 300 nm x-cut LN film on a silicon handle and a BOX thickness of 1.8~$\mu$m. The poling period chosen for this sample is 2.8~$\mu$m. The poling was performed by applying a single, asymmetric, high voltage pulse. The pulse featured a fast ($<1$~ms) ramp up above the coercive field, and was held above the coercive field for a few milliseconds, before being slowly (up to 30 ms) ramped down, to allow stabilization of domains and to prevent spontaneous back switching. More details about the poling setup and pulse can be found in our previous work \cite{Zhao2019}. Subsequently, the sample was investigated with our SH microscopy setup using two objectives with numerical apertures of 0.55 and 0.9. 

The results on the sample are displayed in Fig.~10(b) (objective NA~$=0.55$) and (c) (NA~$=0.9$). Both images show approximately the same sample area, but because some slight repositioning was required after changing the objective, the shown area is not exactly the same. Below each figure is a line plot of the SH intensity through one specific domain highlighted in green. For comparison with theoretical results, the results have been normalized to the intensity observed for un-inverted film, which is shown on the right-hand side of each image. It should be noted that for the measurement with NA~$=0.55$ an excitation power focused on the sample of approximately 5~mW was sufficient, while for NA~$=0.9$ the intensity reaching the sample was reduced to about 2~mW, because the detected single-photon count rate increased by a factor of 4-5 in agreement with the prediction in Sec.~IIIA concerning the use of a higher NA objective. In general, the images with both objective lenses show similar characteristics. The electrodes, which are visible in the bottom and top of the image, show no SH intensity, because the electrode metal (gold) features no second-order nonlinearity in the optical band. The un-inverted film on the right side of each image provides a homogeneous intensity, while the inverted domains are characterized by an area where the intensity has (almost) recovered to 1 normalized units due to the domain inversion. This region is surrounded by a black border, which can be interpreted as the signatures of (possibly slightly angled) domain walls. In both images the targeted domain period of 2.8~$\mu$m is reproduced accurately. The lithographic design of the used poling electrodes feature a slightly lower than 50\% duty cycle, to pre-compensate for the lateral spread of the domains which was observed during growth. This allow us to achieve a close to 50\% duty cycle of the inverted domains. The domain growth started from the bottom electrodes, to which the positive voltage was applied. This is a behavior typical observed during electric field poling of LN, where growth starts from the positive electrodes \cite{Sanna2017}. Figure~10(a) shows a simulated line scan for the given film and numerical aperture parameters for a vertical DW. These calculated line scans can already explain the measured signatures, i.e., width and depth of the measured SH signature. In fact, the FWHM of the DW in the experiment is slightly larger than predicted (about 40~nm). This, for example, can explained  by a slightly angled DW as predicted in Fig.~9. However, what the calculation in Fig.~10(a) cannot explain is that the intensity in the inverted area does not yet recover to the value of the unpoled film. This is, in particular, visible for the higher NA lens due to the increased resolution of the acquired image. This suggests a more complex domain structure, indicating that the domain inversion is not yet complete throughout the film thickness.

Based on the discussion in Sec.~IIIB it should be possible to estimate the inversion depth based on the intensity profile. One possible geometry that might explain our observations is a hexagonal cross section of the poled domain as shown in Fig.~10(d). The domain will spread with approximately the same speed in all three equivalent y-directions and form a hexagonal cross section, with DW's oriented parallel to the y-axis. This is a typical structure for domain structures in congruent LN \cite{Scrymgeour2005a}, and has been observed for x-cut poling  \cite{Berth2009,Gui2009}. If we calculate a line scan for such a structure, we obtain the results shown in the bottom panels of Fig.~10(d). For the calculations given by the black dots, we have assumed that the film is completely inverted at the center of the hexagon ($h_{hex}=300$~nm). In this case, we predict for both numerical apertures an almost complete recovery of the SH intensity to the value in the unpoled film. However, if we assume that there is 10~nm, for example, of un-inverted left below the hexagon shape ($h_{hex}=290$~nm), we calculate signatures (red dots) with good agreement to the experimental data. For comparison, we have plotted the experimental data into the graph (green triangles), which is in good agreement with the calculated profile for $h_{hex}=290$~nm. It should be noted that the assumed domain structure in Fig.~10(d) is hypothetical. The real domain structure may well be more complex. For example, in the image in Fig.~10(c), additional faint black lines closer to the plus electrodes can be seen. This can indicate multiple filaments which are not yet fully merged. Nevertheless, this simulation and experimental results demonstrates that SH microscopy can unambiguously distinguish fully-inverted from non-inverted film, and that the inversion depth may be experimentally determined to the scale of tens of nanometers well below the optical resolution limit. We believe this capability will be an important diagnostic tool in the continued development of poled ferroelectric thin films, especially with smaller poling periods.

\section{Conclusion}

In conclusion, we can explain the observations made with second-harmonic microscopy of poled thin-film lithium niobate. Our model based on a combination of multiple effects, namely reflections, resonant enhancements and phase-(mis)-matching. In particular, the reflectivity of the bottom interfaces below the thin film lead to the detected of SH light being dominated by the co-propagation phase-matching component, rather than the counter-propagation component, which might be na\"ively expected. This explains the surprisingly large intensities of SH light observed from thin-films, as well as the sensitivity of the signal to the relative depth of the domain inversion. Furthermore, we show that SH microscopy is able to unambiguously distinguish between a fully-inverted film and partly-inverted film in depth. The depth of poling can be determined within tens of nanometer resolution. 

There are numerous directions in which this work can be taken further. In our experiment so far, we can explain all observed behaviors without any assumption of an additional crystallographic substructure or symmetry in the range of a DW. However, recent SH investigations show, that ferroelectric DW in LN and other ferroelectrics give rise to new, respectively rotated, tensor elements, due strain \cite{Rusing2018a} or deviations from the ideal Ising-structure of a ferroelectric domain wall \cite{Cherifi-Hertel2017}. For future analysis, these distributions can readily be included in the model. Also, our model assumes a layered sample that extends infinitely into the xy-plane, which is a good assumption for structures larger than the diffraction-limited focus width. However, some periodically-poled devices were fabricated by etching ridge waveguides of micron and sub-micron width in periodically poled LNOI \cite{Wang2018h}. It can be conjectured that 3D diffraction and scattering will have a significant impact on the SH signal and behavior in these structures. Therefore, caution needs to be taken when interpreting signals from such structures, as small changes, e.g. in sidewall angle, can potentially have large impacts on the signal, especially when taking into account the length scales of counter-propagating phase matching or the resonance lengths for the SH light. Taken together, our work suggests that a better understanding of the imaging processes, will enhance the usefulness of SH microscopy as a tool to investigate ferroelectric thin films and improves the technology of poling and the domain wall fabrication for practical applications.

\begin{acknowledgments}
This work was funded through Sandia National Laboratories (SIGMA-NONLin Project) and the National Science Foundation (NSF) (EFMA-1640968).
\end{acknowledgments}


\begin{thebibliography}{62}%
	\makeatletter
	\providecommand \@ifxundefined [1]{%
		\@ifx{#1\undefined}
	}%
	\providecommand \@ifnum [1]{%
		\ifnum #1\expandafter \@firstoftwo
		\else \expandafter \@secondoftwo
		\fi
	}%
	\providecommand \@ifx [1]{%
		\ifx #1\expandafter \@firstoftwo
		\else \expandafter \@secondoftwo
		\fi
	}%
	\providecommand \natexlab [1]{#1}%
	\providecommand \enquote  [1]{``#1''}%
	\providecommand \bibnamefont  [1]{#1}%
	\providecommand \bibfnamefont [1]{#1}%
	\providecommand \citenamefont [1]{#1}%
	\providecommand \href@noop [0]{\@secondoftwo}%
	\providecommand \href [0]{\begingroup \@sanitize@url \@href}%
	\providecommand \@href[1]{\@@startlink{#1}\@@href}%
	\providecommand \@@href[1]{\endgroup#1\@@endlink}%
	\providecommand \@sanitize@url [0]{\catcode `\\12\catcode `\$12\catcode
		`\&12\catcode `\#12\catcode `\^12\catcode `\_12\catcode `\%12\relax}%
	\providecommand \@@startlink[1]{}%
	\providecommand \@@endlink[0]{}%
	\providecommand \url  [0]{\begingroup\@sanitize@url \@url }%
	\providecommand \@url [1]{\endgroup\@href {#1}{\urlprefix }}%
	\providecommand \urlprefix  [0]{URL }%
	\providecommand \Eprint [0]{\href }%
	\providecommand \doibase [0]{http://dx.doi.org/}%
	\providecommand \selectlanguage [0]{\@gobble}%
	\providecommand \bibinfo  [0]{\@secondoftwo}%
	\providecommand \bibfield  [0]{\@secondoftwo}%
	\providecommand \translation [1]{[#1]}%
	\providecommand \BibitemOpen [0]{}%
	\providecommand \bibitemStop [0]{}%
	\providecommand \bibitemNoStop [0]{.\EOS\space}%
	\providecommand \EOS [0]{\spacefactor3000\relax}%
	\providecommand \BibitemShut  [1]{\csname bibitem#1\endcsname}%
	\let\auto@bib@innerbib\@empty
	%</preamble>
	\bibitem [{\citenamefont {Berth}\ \emph {et~al.}(2007)\citenamefont {Berth},
		\citenamefont {Quiring}, \citenamefont {Sohler},\ and\ \citenamefont
		{Zrenner}}]{Berth2007}%
	\BibitemOpen
	\bibfield  {author} {\bibinfo {author} {\bibfnamefont {G.}~\bibnamefont
			{Berth}}, \bibinfo {author} {\bibfnamefont {V.}~\bibnamefont {Quiring}},
		\bibinfo {author} {\bibfnamefont {W.}~\bibnamefont {Sohler}}, \ and\ \bibinfo
		{author} {\bibfnamefont {A.}~\bibnamefont {Zrenner}},\ }\href {\doibase
		10.1080/00150190701358159} {\bibfield  {journal} {\bibinfo  {journal}
			{Ferroelectrics}\ }\textbf {\bibinfo {volume} {352}},\ \bibinfo {pages} {78}
		(\bibinfo {year} {2007})}\BibitemShut {NoStop}%
	\bibitem [{\citenamefont {Berth}\ \emph {et~al.}(2009)\citenamefont {Berth},
		\citenamefont {Wiedemeier}, \citenamefont {H{\"{u}}sch}, \citenamefont {Gui},
		\citenamefont {Hu}, \citenamefont {Sohler},\ and\ \citenamefont
		{Zrenner}}]{Berth2009}%
	\BibitemOpen
	\bibfield  {author} {\bibinfo {author} {\bibfnamefont {G.}~\bibnamefont
			{Berth}}, \bibinfo {author} {\bibfnamefont {V.}~\bibnamefont {Wiedemeier}},
		\bibinfo {author} {\bibfnamefont {K.~P.}\ \bibnamefont {H{\"{u}}sch}},
		\bibinfo {author} {\bibfnamefont {L.}~\bibnamefont {Gui}}, \bibinfo {author}
		{\bibfnamefont {H.}~\bibnamefont {Hu}}, \bibinfo {author} {\bibfnamefont
			{W.}~\bibnamefont {Sohler}}, \ and\ \bibinfo {author} {\bibfnamefont
			{A.}~\bibnamefont {Zrenner}},\ }\href {\doibase 10.1080/00150190902993267}
	{\bibfield  {journal} {\bibinfo  {journal} {Ferroelectrics}\ }\textbf
		{\bibinfo {volume} {389}},\ \bibinfo {pages} {132} (\bibinfo {year}
		{2009})}\BibitemShut {NoStop}%
	\bibitem [{\citenamefont {Mackwitz}\ \emph {et~al.}(2016)\citenamefont
		{Mackwitz}, \citenamefont {R{\"{u}}sing}, \citenamefont {Berth},
		\citenamefont {Widhalm}, \citenamefont {M{\"{u}}ller},\ and\ \citenamefont
		{Zrenner}}]{Mackwitz2016}%
	\BibitemOpen
	\bibfield  {author} {\bibinfo {author} {\bibfnamefont {P.}~\bibnamefont
			{Mackwitz}}, \bibinfo {author} {\bibfnamefont {M.}~\bibnamefont
			{R{\"{u}}sing}}, \bibinfo {author} {\bibfnamefont {G.}~\bibnamefont {Berth}},
		\bibinfo {author} {\bibfnamefont {A.}~\bibnamefont {Widhalm}}, \bibinfo
		{author} {\bibfnamefont {K.}~\bibnamefont {M{\"{u}}ller}}, \ and\ \bibinfo
		{author} {\bibfnamefont {A.}~\bibnamefont {Zrenner}},\ }\href {\doibase
		10.1063/1.4946010} {\bibfield  {journal} {\bibinfo  {journal} {Applied
				Physics Letters}\ }\textbf {\bibinfo {volume} {108}},\ \bibinfo {pages}
		{152902} (\bibinfo {year} {2016})}\BibitemShut {NoStop}%
	\bibitem [{\citenamefont {Fl{\"{o}}rsheimer}\ \emph {et~al.}(1998)\citenamefont
		{Fl{\"{o}}rsheimer}, \citenamefont {Paschotta}, \citenamefont {Kubitscheck},
		\citenamefont {Brillert}, \citenamefont {Hofmann}, \citenamefont {Heuer},
		\citenamefont {Schreiber}, \citenamefont {Verbeek}, \citenamefont {Sohler},\
		and\ \citenamefont {Fuchs}}]{Florsheimer1998}%
	\BibitemOpen
	\bibfield  {author} {\bibinfo {author} {\bibfnamefont {M.}~\bibnamefont
			{Fl{\"{o}}rsheimer}}, \bibinfo {author} {\bibfnamefont {R.}~\bibnamefont
			{Paschotta}}, \bibinfo {author} {\bibfnamefont {U.}~\bibnamefont
			{Kubitscheck}}, \bibinfo {author} {\bibfnamefont {C.}~\bibnamefont
			{Brillert}}, \bibinfo {author} {\bibfnamefont {D.}~\bibnamefont {Hofmann}},
		\bibinfo {author} {\bibfnamefont {L.}~\bibnamefont {Heuer}}, \bibinfo
		{author} {\bibfnamefont {G.}~\bibnamefont {Schreiber}}, \bibinfo {author}
		{\bibfnamefont {C.}~\bibnamefont {Verbeek}}, \bibinfo {author} {\bibfnamefont
			{W.}~\bibnamefont {Sohler}}, \ and\ \bibinfo {author} {\bibfnamefont
			{H.}~\bibnamefont {Fuchs}},\ }\href {\doibase 10.1007/s003400050552}
	{\bibfield  {journal} {\bibinfo  {journal} {Applied Physics B: Lasers and
				Optics}\ }\textbf {\bibinfo {volume} {67}},\ \bibinfo {pages} {593} (\bibinfo
		{year} {1998})}\BibitemShut {NoStop}%
	\bibitem [{\citenamefont {Bozhevolnyi}\ \emph {et~al.}(1998)\citenamefont
		{Bozhevolnyi}, \citenamefont {Hvam}, \citenamefont {Pedersen}, \citenamefont
		{Laurell}, \citenamefont {Karlsson}, \citenamefont {Skettrup},\ and\
		\citenamefont {Belmonte}}]{Bozhevolnyi1998}%
	\BibitemOpen
	\bibfield  {author} {\bibinfo {author} {\bibfnamefont {S.~I.}\ \bibnamefont
			{Bozhevolnyi}}, \bibinfo {author} {\bibfnamefont {J.~M.}\ \bibnamefont
			{Hvam}}, \bibinfo {author} {\bibfnamefont {K.}~\bibnamefont {Pedersen}},
		\bibinfo {author} {\bibfnamefont {F.}~\bibnamefont {Laurell}}, \bibinfo
		{author} {\bibfnamefont {H.}~\bibnamefont {Karlsson}}, \bibinfo {author}
		{\bibfnamefont {T.}~\bibnamefont {Skettrup}}, \ and\ \bibinfo {author}
		{\bibfnamefont {M.}~\bibnamefont {Belmonte}},\ }\href {\doibase
		10.1063/1.122291} {\bibfield  {journal} {\bibinfo  {journal} {Applied Physics
				Letters}\ }\textbf {\bibinfo {volume} {73}},\ \bibinfo {pages} {1814}
		(\bibinfo {year} {1998})}\BibitemShut {NoStop}%
	\bibitem [{\citenamefont {Cherifi-Hertel}\ \emph {et~al.}(2017)\citenamefont
		{Cherifi-Hertel}, \citenamefont {Bulou}, \citenamefont {Hertel},
		\citenamefont {Taupier}, \citenamefont {Dorkenoo}, \citenamefont {Andreas},
		\citenamefont {Guyonnet}, \citenamefont {Gaponenko}, \citenamefont {Gallo},\
		and\ \citenamefont {Paruch}}]{Cherifi-Hertel2017}%
	\BibitemOpen
	\bibfield  {author} {\bibinfo {author} {\bibfnamefont {S.}~\bibnamefont
			{Cherifi-Hertel}}, \bibinfo {author} {\bibfnamefont {H.}~\bibnamefont
			{Bulou}}, \bibinfo {author} {\bibfnamefont {R.}~\bibnamefont {Hertel}},
		\bibinfo {author} {\bibfnamefont {G.}~\bibnamefont {Taupier}}, \bibinfo
		{author} {\bibfnamefont {K.~D.~H.}\ \bibnamefont {Dorkenoo}}, \bibinfo
		{author} {\bibfnamefont {C.}~\bibnamefont {Andreas}}, \bibinfo {author}
		{\bibfnamefont {J.}~\bibnamefont {Guyonnet}}, \bibinfo {author}
		{\bibfnamefont {I.}~\bibnamefont {Gaponenko}}, \bibinfo {author}
		{\bibfnamefont {K.}~\bibnamefont {Gallo}}, \ and\ \bibinfo {author}
		{\bibfnamefont {P.}~\bibnamefont {Paruch}},\ }\href {\doibase
		10.1038/ncomms15768} {\bibfield  {journal} {\bibinfo  {journal} {Nature
				Communications}\ }\textbf {\bibinfo {volume} {8}},\ \bibinfo {pages} {15768}
		(\bibinfo {year} {2017})}\BibitemShut {NoStop}%
	\bibitem [{\citenamefont {Huang}\ \emph {et~al.}(2017)\citenamefont {Huang},
		\citenamefont {Wei}, \citenamefont {Wang}, \citenamefont {Zhu}, \citenamefont
		{Zhang}, \citenamefont {Hu}, \citenamefont {Zhu},\ and\ \citenamefont
		{Xiao}}]{Huang2017}%
	\BibitemOpen
	\bibfield  {author} {\bibinfo {author} {\bibfnamefont {X.}~\bibnamefont
			{Huang}}, \bibinfo {author} {\bibfnamefont {D.}~\bibnamefont {Wei}}, \bibinfo
		{author} {\bibfnamefont {Y.}~\bibnamefont {Wang}}, \bibinfo {author}
		{\bibfnamefont {Y.}~\bibnamefont {Zhu}}, \bibinfo {author} {\bibfnamefont
			{Y.}~\bibnamefont {Zhang}}, \bibinfo {author} {\bibfnamefont {X.~P.}\
			\bibnamefont {Hu}}, \bibinfo {author} {\bibfnamefont {S.~N.}\ \bibnamefont
			{Zhu}}, \ and\ \bibinfo {author} {\bibfnamefont {M.}~\bibnamefont {Xiao}},\
	}\href {\doibase 10.1088/1361-6463/aa9258} {\bibfield  {journal} {\bibinfo
			{journal} {Journal of Physics D: Applied Physics}\ }\textbf {\bibinfo
			{volume} {50}},\ \bibinfo {pages} {485105} (\bibinfo {year}
		{2017})}\BibitemShut {NoStop}%
	\bibitem [{\citenamefont {Zhao}, \citenamefont {R{\"{u}}sing},\ and\
		\citenamefont {Mookherjea}(2019)}]{Zhao2019}%
	\BibitemOpen
	\bibfield  {author} {\bibinfo {author} {\bibfnamefont {J.}~\bibnamefont
			{Zhao}}, \bibinfo {author} {\bibfnamefont {M.}~\bibnamefont {R{\"{u}}sing}},
		\ and\ \bibinfo {author} {\bibfnamefont {S.}~\bibnamefont {Mookherjea}},\
	}\href {\doibase 10.1364/OE.27.012025} {\bibfield  {journal} {\bibinfo
			{journal} {Optics Express}\ }\textbf {\bibinfo {volume} {27}},\ \bibinfo
		{pages} {12025} (\bibinfo {year} {2019})}\BibitemShut {NoStop}%
	\bibitem [{\citenamefont {Kaneshiro}, \citenamefont {Uesu},\ and\ \citenamefont
		{Fukui}(2010)}]{Kaneshiro2010}%
	\BibitemOpen
	\bibfield  {author} {\bibinfo {author} {\bibfnamefont {J.}~\bibnamefont
			{Kaneshiro}}, \bibinfo {author} {\bibfnamefont {Y.}~\bibnamefont {Uesu}}, \
		and\ \bibinfo {author} {\bibfnamefont {T.}~\bibnamefont {Fukui}},\ }\href
	{\doibase 10.1364/josab.27.000888} {\bibfield  {journal} {\bibinfo  {journal}
			{Journal of the Optical Society of America B}\ }\textbf {\bibinfo {volume}
			{27}},\ \bibinfo {pages} {888} (\bibinfo {year} {2010})}\BibitemShut
	{NoStop}%
	\bibitem [{\citenamefont {Kurimura}\ and\ \citenamefont
		{Uesu}(1997)}]{Kurimura1997}%
	\BibitemOpen
	\bibfield  {author} {\bibinfo {author} {\bibfnamefont {S.}~\bibnamefont
			{Kurimura}}\ and\ \bibinfo {author} {\bibfnamefont {Y.}~\bibnamefont
			{Uesu}},\ }\href {\doibase 10.1063/1.364121} {\bibfield  {journal} {\bibinfo
			{journal} {Journal of Applied Physics}\ }\textbf {\bibinfo {volume} {81}},\
		\bibinfo {pages} {369} (\bibinfo {year} {1997})}\BibitemShut {NoStop}%
	\bibitem [{\citenamefont {Spychala}\ \emph {et~al.}(2017)\citenamefont
		{Spychala}, \citenamefont {Berth}, \citenamefont {Widhalm}, \citenamefont
		{R{\"{u}}sing}, \citenamefont {Wang}, \citenamefont {Sanna},\ and\
		\citenamefont {Zrenner}}]{Spychala2017}%
	\BibitemOpen
	\bibfield  {author} {\bibinfo {author} {\bibfnamefont {K.~J.}\ \bibnamefont
			{Spychala}}, \bibinfo {author} {\bibfnamefont {G.}~\bibnamefont {Berth}},
		\bibinfo {author} {\bibfnamefont {A.}~\bibnamefont {Widhalm}}, \bibinfo
		{author} {\bibfnamefont {M.}~\bibnamefont {R{\"{u}}sing}}, \bibinfo {author}
		{\bibfnamefont {L.}~\bibnamefont {Wang}}, \bibinfo {author} {\bibfnamefont
			{S.}~\bibnamefont {Sanna}}, \ and\ \bibinfo {author} {\bibfnamefont
			{A.}~\bibnamefont {Zrenner}},\ }\href {\doibase 10.1364/OE.25.021444}
	{\bibfield  {journal} {\bibinfo  {journal} {Optics Express}\ }\textbf
		{\bibinfo {volume} {25}},\ \bibinfo {pages} {21444} (\bibinfo {year}
		{2017})}\BibitemShut {NoStop}%
	\bibitem [{\citenamefont {Wu}\ \emph {et~al.}(2012)\citenamefont {Wu},
		\citenamefont {Horibe}, \citenamefont {Lee}, \citenamefont {Cheong},\ and\
		\citenamefont {Guest}}]{Wu2012}%
	\BibitemOpen
	\bibfield  {author} {\bibinfo {author} {\bibfnamefont {W.}~\bibnamefont
			{Wu}}, \bibinfo {author} {\bibfnamefont {Y.}~\bibnamefont {Horibe}}, \bibinfo
		{author} {\bibfnamefont {N.}~\bibnamefont {Lee}}, \bibinfo {author}
		{\bibfnamefont {S.-W.}\ \bibnamefont {Cheong}}, \ and\ \bibinfo {author}
		{\bibfnamefont {J.~R.}\ \bibnamefont {Guest}},\ }\href {\doibase
		10.1103/PhysRevLett.108.077203} {\bibfield  {journal} {\bibinfo  {journal}
			{Physical Review Letters}\ }\textbf {\bibinfo {volume} {108}},\ \bibinfo
		{pages} {077203} (\bibinfo {year} {2012})}\BibitemShut {NoStop}%
	\bibitem [{\citenamefont {Kumagai}\ and\ \citenamefont
		{Spaldin}(2013)}]{Kumagai2013}%
	\BibitemOpen
	\bibfield  {author} {\bibinfo {author} {\bibfnamefont {Y.}~\bibnamefont
			{Kumagai}}\ and\ \bibinfo {author} {\bibfnamefont {N.~A.}\ \bibnamefont
			{Spaldin}},\ }\href {\doibase 10.1038/ncomms2545} {\bibfield  {journal}
		{\bibinfo  {journal} {Nature Communications}\ }\textbf {\bibinfo {volume}
			{4}},\ \bibinfo {pages} {1540} (\bibinfo {year} {2013})}\BibitemShut
	{NoStop}%
	\bibitem [{\citenamefont {Ma}\ \emph {et~al.}(2018)\citenamefont {Ma},
		\citenamefont {Ma}, \citenamefont {Zhang}, \citenamefont {Peng},
		\citenamefont {Wang}, \citenamefont {Liu}, \citenamefont {Wang},
		\citenamefont {Li}, \citenamefont {Chen}, \citenamefont {Cheng},
		\citenamefont {Gao}, \citenamefont {Gu}, \citenamefont {Chen}, \citenamefont
		{Yu}, \citenamefont {Zhang},\ and\ \citenamefont {Nan}}]{Ma2018}%
	\BibitemOpen
	\bibfield  {author} {\bibinfo {author} {\bibfnamefont {J.}~\bibnamefont
			{Ma}}, \bibinfo {author} {\bibfnamefont {J.}~\bibnamefont {Ma}}, \bibinfo
		{author} {\bibfnamefont {Q.}~\bibnamefont {Zhang}}, \bibinfo {author}
		{\bibfnamefont {R.}~\bibnamefont {Peng}}, \bibinfo {author} {\bibfnamefont
			{J.}~\bibnamefont {Wang}}, \bibinfo {author} {\bibfnamefont {C.}~\bibnamefont
			{Liu}}, \bibinfo {author} {\bibfnamefont {M.}~\bibnamefont {Wang}}, \bibinfo
		{author} {\bibfnamefont {N.}~\bibnamefont {Li}}, \bibinfo {author}
		{\bibfnamefont {M.}~\bibnamefont {Chen}}, \bibinfo {author} {\bibfnamefont
			{X.}~\bibnamefont {Cheng}}, \bibinfo {author} {\bibfnamefont
			{P.}~\bibnamefont {Gao}}, \bibinfo {author} {\bibfnamefont {L.}~\bibnamefont
			{Gu}}, \bibinfo {author} {\bibfnamefont {L.-Q.}\ \bibnamefont {Chen}},
		\bibinfo {author} {\bibfnamefont {P.}~\bibnamefont {Yu}}, \bibinfo {author}
		{\bibfnamefont {J.}~\bibnamefont {Zhang}}, \ and\ \bibinfo {author}
		{\bibfnamefont {C.-W.}\ \bibnamefont {Nan}},\ }\href {\doibase
		10.1038/s41565-018-0204-1} {\bibfield  {journal} {\bibinfo  {journal} {Nature
				Nanotechnology}\ }\textbf {\bibinfo {volume} {13}},\ \bibinfo {pages} {947}
		(\bibinfo {year} {2018})}\BibitemShut {NoStop}%
	\bibitem [{\citenamefont {Nahas}\ \emph {et~al.}(2015)\citenamefont {Nahas},
		\citenamefont {Prokhorenko}, \citenamefont {Louis}, \citenamefont {Gui},
		\citenamefont {Kornev},\ and\ \citenamefont {Bellaiche}}]{Nahas2015}%
	\BibitemOpen
	\bibfield  {author} {\bibinfo {author} {\bibfnamefont {Y.}~\bibnamefont
			{Nahas}}, \bibinfo {author} {\bibfnamefont {S.}~\bibnamefont {Prokhorenko}},
		\bibinfo {author} {\bibfnamefont {L.}~\bibnamefont {Louis}}, \bibinfo
		{author} {\bibfnamefont {Z.}~\bibnamefont {Gui}}, \bibinfo {author}
		{\bibfnamefont {I.}~\bibnamefont {Kornev}}, \ and\ \bibinfo {author}
		{\bibfnamefont {L.}~\bibnamefont {Bellaiche}},\ }\href {\doibase
		10.1038/ncomms9542} {\bibfield  {journal} {\bibinfo  {journal} {Nature
				Communications}\ }\textbf {\bibinfo {volume} {6}},\ \bibinfo {pages} {8542}
		(\bibinfo {year} {2015})}\BibitemShut {NoStop}%
	\bibitem [{\citenamefont {Sharma}\ \emph {et~al.}(2017)\citenamefont {Sharma},
		\citenamefont {Zhang}, \citenamefont {Sando}, \citenamefont {Lei},
		\citenamefont {Liu}, \citenamefont {Li}, \citenamefont {Nagarajan},\ and\
		\citenamefont {Seidel}}]{Sharma2017}%
	\BibitemOpen
	\bibfield  {author} {\bibinfo {author} {\bibfnamefont {P.}~\bibnamefont
			{Sharma}}, \bibinfo {author} {\bibfnamefont {Q.}~\bibnamefont {Zhang}},
		\bibinfo {author} {\bibfnamefont {D.}~\bibnamefont {Sando}}, \bibinfo
		{author} {\bibfnamefont {C.~H.}\ \bibnamefont {Lei}}, \bibinfo {author}
		{\bibfnamefont {Y.}~\bibnamefont {Liu}}, \bibinfo {author} {\bibfnamefont
			{J.}~\bibnamefont {Li}}, \bibinfo {author} {\bibfnamefont {V.}~\bibnamefont
			{Nagarajan}}, \ and\ \bibinfo {author} {\bibfnamefont {J.}~\bibnamefont
			{Seidel}},\ }\href {\doibase 10.1126/sciadv.1700512} {\bibfield  {journal}
		{\bibinfo  {journal} {Science Advances}\ }\textbf {\bibinfo {volume} {3}},\
		\bibinfo {pages} {1} (\bibinfo {year} {2017})}\BibitemShut {NoStop}%
	\bibitem [{\citenamefont {Godau}\ \emph {et~al.}(2017)\citenamefont {Godau},
		\citenamefont {K{\"{a}}mpfe}, \citenamefont {Thiessen}, \citenamefont {Eng},\
		and\ \citenamefont {Hau{\ss}mann}}]{Godau2017}%
	\BibitemOpen
	\bibfield  {author} {\bibinfo {author} {\bibfnamefont {C.}~\bibnamefont
			{Godau}}, \bibinfo {author} {\bibfnamefont {T.}~\bibnamefont {K{\"{a}}mpfe}},
		\bibinfo {author} {\bibfnamefont {A.}~\bibnamefont {Thiessen}}, \bibinfo
		{author} {\bibfnamefont {L.~M.}\ \bibnamefont {Eng}}, \ and\ \bibinfo
		{author} {\bibfnamefont {A.}~\bibnamefont {Hau{\ss}mann}},\ }\href {\doibase
		10.1021/acsnano.7b01199} {\bibfield  {journal} {\bibinfo  {journal} {ACS
				Nano}\ }\textbf {\bibinfo {volume} {11}},\ \bibinfo {pages} {4816} (\bibinfo
		{year} {2017})}\BibitemShut {NoStop}%
	\bibitem [{\citenamefont {Schr{\"{o}}der}\ \emph {et~al.}(2012)\citenamefont
		{Schr{\"{o}}der}, \citenamefont {Hau{\ss}mann}, \citenamefont {Thiessen},
		\citenamefont {Soergel}, \citenamefont {Woike},\ and\ \citenamefont
		{Eng}}]{Schroder2012}%
	\BibitemOpen
	\bibfield  {author} {\bibinfo {author} {\bibfnamefont {M.}~\bibnamefont
			{Schr{\"{o}}der}}, \bibinfo {author} {\bibfnamefont {A.}~\bibnamefont
			{Hau{\ss}mann}}, \bibinfo {author} {\bibfnamefont {A.}~\bibnamefont
			{Thiessen}}, \bibinfo {author} {\bibfnamefont {E.}~\bibnamefont {Soergel}},
		\bibinfo {author} {\bibfnamefont {T.}~\bibnamefont {Woike}}, \ and\ \bibinfo
		{author} {\bibfnamefont {L.~M.}\ \bibnamefont {Eng}},\ }\href {\doibase
		10.1002/adfm.201201174} {\bibfield  {journal} {\bibinfo  {journal} {Advanced
				Functional Materials}\ }\textbf {\bibinfo {volume} {22}},\ \bibinfo {pages}
		{3936} (\bibinfo {year} {2012})}\BibitemShut {NoStop}%
	\bibitem [{\citenamefont {Wolba}\ \emph {et~al.}(2018)\citenamefont {Wolba},
		\citenamefont {Seidel}, \citenamefont {Cazorla}, \citenamefont {Godau},
		\citenamefont {Hau{\ss}mann},\ and\ \citenamefont {Eng}}]{Wolba2018}%
	\BibitemOpen
	\bibfield  {author} {\bibinfo {author} {\bibfnamefont {B.}~\bibnamefont
			{Wolba}}, \bibinfo {author} {\bibfnamefont {J.}~\bibnamefont {Seidel}},
		\bibinfo {author} {\bibfnamefont {C.}~\bibnamefont {Cazorla}}, \bibinfo
		{author} {\bibfnamefont {C.}~\bibnamefont {Godau}}, \bibinfo {author}
		{\bibfnamefont {A.}~\bibnamefont {Hau{\ss}mann}}, \ and\ \bibinfo {author}
		{\bibfnamefont {L.~M.}\ \bibnamefont {Eng}},\ }\href {\doibase
		10.1002/aelm.201700242} {\bibfield  {journal} {\bibinfo  {journal} {Advanced
				Electronic Materials}\ }\textbf {\bibinfo {volume} {4}},\ \bibinfo {pages}
		{1700242} (\bibinfo {year} {2018})}\BibitemShut {NoStop}%
	\bibitem [{\citenamefont {Wang}\ \emph {et~al.}(2018)\citenamefont {Wang},
		\citenamefont {Langrock}, \citenamefont {Marandi}, \citenamefont {Jankowski},
		\citenamefont {Zhang}, \citenamefont {Desiatov}, \citenamefont {Fejer},\ and\
		\citenamefont {Lon{\v{c}}ar}}]{Wang2018h}%
	\BibitemOpen
	\bibfield  {author} {\bibinfo {author} {\bibfnamefont {C.}~\bibnamefont
			{Wang}}, \bibinfo {author} {\bibfnamefont {C.}~\bibnamefont {Langrock}},
		\bibinfo {author} {\bibfnamefont {A.}~\bibnamefont {Marandi}}, \bibinfo
		{author} {\bibfnamefont {M.}~\bibnamefont {Jankowski}}, \bibinfo {author}
		{\bibfnamefont {M.}~\bibnamefont {Zhang}}, \bibinfo {author} {\bibfnamefont
			{B.}~\bibnamefont {Desiatov}}, \bibinfo {author} {\bibfnamefont {M.~M.}\
			\bibnamefont {Fejer}}, \ and\ \bibinfo {author} {\bibfnamefont
			{M.}~\bibnamefont {Lon{\v{c}}ar}},\ }\href {\doibase 10.1364/OPTICA.5.001438}
	{\bibfield  {journal} {\bibinfo  {journal} {Optica}\ }\textbf {\bibinfo
			{volume} {5}},\ \bibinfo {pages} {1438} (\bibinfo {year} {2018})}\BibitemShut
	{NoStop}%
	\bibitem [{\citenamefont {Chang}\ \emph {et~al.}(2016)\citenamefont {Chang},
		\citenamefont {Li}, \citenamefont {Volet}, \citenamefont {Wang},
		\citenamefont {Peters},\ and\ \citenamefont {Bowers}}]{Chang2016}%
	\BibitemOpen
	\bibfield  {author} {\bibinfo {author} {\bibfnamefont {L.}~\bibnamefont
			{Chang}}, \bibinfo {author} {\bibfnamefont {Y.}~\bibnamefont {Li}}, \bibinfo
		{author} {\bibfnamefont {N.}~\bibnamefont {Volet}}, \bibinfo {author}
		{\bibfnamefont {L.}~\bibnamefont {Wang}}, \bibinfo {author} {\bibfnamefont
			{J.}~\bibnamefont {Peters}}, \ and\ \bibinfo {author} {\bibfnamefont {J.~E.}\
			\bibnamefont {Bowers}},\ }\href {\doibase 10.1364/OPTICA.3.000531} {\bibfield
		{journal} {\bibinfo  {journal} {Optica}\ }\textbf {\bibinfo {volume} {3}},\
		\bibinfo {pages} {531} (\bibinfo {year} {2016})}\BibitemShut {NoStop}%
	\bibitem [{\citenamefont {Weigel}\ and\ \citenamefont
		{Mookherjea}(2018)}]{Weigel2018}%
	\BibitemOpen
	\bibfield  {author} {\bibinfo {author} {\bibfnamefont {P.}~\bibnamefont
			{Weigel}}\ and\ \bibinfo {author} {\bibfnamefont {S.}~\bibnamefont
			{Mookherjea}},\ }\href {\doibase 10.1364/JOSAB.35.000593} {\bibfield
		{journal} {\bibinfo  {journal} {Journal of the Optical Society of America B:
				Optical Physics}\ }\textbf {\bibinfo {volume} {35}},\ \bibinfo {pages} {593}
		(\bibinfo {year} {2018})}\BibitemShut {NoStop}%
	\bibitem [{\citenamefont {Allgaier}\ \emph {et~al.}(2017)\citenamefont
		{Allgaier}, \citenamefont {Ansari}, \citenamefont {Sansoni}, \citenamefont
		{Eigner}, \citenamefont {Quiring}, \citenamefont {Ricken}, \citenamefont
		{Harder}, \citenamefont {Brecht},\ and\ \citenamefont
		{Silberhorn}}]{Allgaier2017}%
	\BibitemOpen
	\bibfield  {author} {\bibinfo {author} {\bibfnamefont {M.}~\bibnamefont
			{Allgaier}}, \bibinfo {author} {\bibfnamefont {V.}~\bibnamefont {Ansari}},
		\bibinfo {author} {\bibfnamefont {L.}~\bibnamefont {Sansoni}}, \bibinfo
		{author} {\bibfnamefont {C.}~\bibnamefont {Eigner}}, \bibinfo {author}
		{\bibfnamefont {V.}~\bibnamefont {Quiring}}, \bibinfo {author} {\bibfnamefont
			{R.}~\bibnamefont {Ricken}}, \bibinfo {author} {\bibfnamefont
			{G.}~\bibnamefont {Harder}}, \bibinfo {author} {\bibfnamefont
			{B.}~\bibnamefont {Brecht}}, \ and\ \bibinfo {author} {\bibfnamefont
			{C.}~\bibnamefont {Silberhorn}},\ }\href {\doibase 10.1038/ncomms14288}
	{\bibfield  {journal} {\bibinfo  {journal} {Nature Communications}\ }\textbf
		{\bibinfo {volume} {8}},\ \bibinfo {pages} {1} (\bibinfo {year} {2017})},\
	\Eprint {http://arxiv.org/abs/1610.08326} {arXiv:1610.08326} \BibitemShut
	{NoStop}%
	\bibitem [{\citenamefont {Sharapova}\ \emph {et~al.}(2017)\citenamefont
		{Sharapova}, \citenamefont {Luo}, \citenamefont {Herrmann}, \citenamefont
		{Reichelt}, \citenamefont {Meier},\ and\ \citenamefont
		{Silberhorn}}]{Sharapova2017}%
	\BibitemOpen
	\bibfield  {author} {\bibinfo {author} {\bibfnamefont {P.~R.}\ \bibnamefont
			{Sharapova}}, \bibinfo {author} {\bibfnamefont {K.~H.}\ \bibnamefont {Luo}},
		\bibinfo {author} {\bibfnamefont {H.}~\bibnamefont {Herrmann}}, \bibinfo
		{author} {\bibfnamefont {M.}~\bibnamefont {Reichelt}}, \bibinfo {author}
		{\bibfnamefont {T.}~\bibnamefont {Meier}}, \ and\ \bibinfo {author}
		{\bibfnamefont {C.}~\bibnamefont {Silberhorn}},\ }\href {\doibase
		10.1088/1367-2630/aa9033} {\bibfield  {journal} {\bibinfo  {journal} {New
				Journal of Physics}\ }\textbf {\bibinfo {volume} {19}},\ \bibinfo {pages}
		{123009} (\bibinfo {year} {2017})},\ \Eprint
	{http://arxiv.org/abs/1704.03769} {arXiv:1704.03769} \BibitemShut {NoStop}%
	\bibitem [{\citenamefont {Imeshev}\ \emph {et~al.}(2001)\citenamefont
		{Imeshev}, \citenamefont {Arbore}, \citenamefont {Fejer}, \citenamefont
		{Galvanauskas}, \citenamefont {Fermann},\ and\ \citenamefont
		{Harter}}]{Imeshev2001}%
	\BibitemOpen
	\bibfield  {author} {\bibinfo {author} {\bibfnamefont {G.}~\bibnamefont
			{Imeshev}}, \bibinfo {author} {\bibfnamefont {M.~A.}\ \bibnamefont {Arbore}},
		\bibinfo {author} {\bibfnamefont {M.~M.}\ \bibnamefont {Fejer}}, \bibinfo
		{author} {\bibfnamefont {A.}~\bibnamefont {Galvanauskas}}, \bibinfo {author}
		{\bibfnamefont {M.}~\bibnamefont {Fermann}}, \ and\ \bibinfo {author}
		{\bibfnamefont {D.}~\bibnamefont {Harter}},\ }\href {\doibase
		10.1364/JOSAB.18.000121} {\bibfield  {journal} {\bibinfo  {journal} {Journal
				of the Optical Society of America B}\ }\textbf {\bibinfo {volume} {18}},\
		\bibinfo {pages} {121} (\bibinfo {year} {2001})}\BibitemShut {NoStop}%
	\bibitem [{\citenamefont {Garmire}(2013)}]{Garmire2013}%
	\BibitemOpen
	\bibfield  {author} {\bibinfo {author} {\bibfnamefont {E.}~\bibnamefont
			{Garmire}},\ }\href {\doibase 10.1364/OE.21.030532} {\bibfield  {journal}
		{\bibinfo  {journal} {Optics Express}\ }\textbf {\bibinfo {volume} {21}},\
		\bibinfo {pages} {30532} (\bibinfo {year} {2013})}\BibitemShut {NoStop}%
	\bibitem [{\citenamefont {Fernandez-Gonzalvo}\ \emph
		{et~al.}(2013)\citenamefont {Fernandez-Gonzalvo}, \citenamefont {Corrielli},
		\citenamefont {Albrecht}, \citenamefont {Grimau}, \citenamefont {Cristiani},\
		and\ \citenamefont {de~Riedmatten}}]{Fernandez-Gonzalvo2013}%
	\BibitemOpen
	\bibfield  {author} {\bibinfo {author} {\bibfnamefont {X.}~\bibnamefont
			{Fernandez-Gonzalvo}}, \bibinfo {author} {\bibfnamefont {G.}~\bibnamefont
			{Corrielli}}, \bibinfo {author} {\bibfnamefont {B.}~\bibnamefont {Albrecht}},
		\bibinfo {author} {\bibfnamefont {M.}~\bibnamefont {Grimau}}, \bibinfo
		{author} {\bibfnamefont {M.}~\bibnamefont {Cristiani}}, \ and\ \bibinfo
		{author} {\bibfnamefont {H.}~\bibnamefont {de~Riedmatten}},\ }\href {\doibase
		10.1364/OE.21.019473} {\bibfield  {journal} {\bibinfo  {journal} {Optics
				Express}\ }\textbf {\bibinfo {volume} {21}},\ \bibinfo {pages} {19473}
		(\bibinfo {year} {2013})}\BibitemShut {NoStop}%
	\bibitem [{\citenamefont {Maring}\ \emph {et~al.}(2018)\citenamefont {Maring},
		\citenamefont {Lago-Rivera}, \citenamefont {Lenhard}, \citenamefont
		{Heinze},\ and\ \citenamefont {de~Riedmatten}}]{Maring2018}%
	\BibitemOpen
	\bibfield  {author} {\bibinfo {author} {\bibfnamefont {N.}~\bibnamefont
			{Maring}}, \bibinfo {author} {\bibfnamefont {D.}~\bibnamefont {Lago-Rivera}},
		\bibinfo {author} {\bibfnamefont {A.}~\bibnamefont {Lenhard}}, \bibinfo
		{author} {\bibfnamefont {G.}~\bibnamefont {Heinze}}, \ and\ \bibinfo {author}
		{\bibfnamefont {H.}~\bibnamefont {de~Riedmatten}},\ }\href {\doibase
		10.1364/OPTICA.5.000507} {\bibfield  {journal} {\bibinfo  {journal} {Optica}\
		}\textbf {\bibinfo {volume} {5}},\ \bibinfo {pages} {507} (\bibinfo {year}
		{2018})},\ \Eprint {http://arxiv.org/abs/1801.03727} {arXiv:1801.03727}
	\BibitemShut {NoStop}%
	\bibitem [{\citenamefont {Weis}\ and\ \citenamefont
		{Gaylord}(1985)}]{Weis1985}%
	\BibitemOpen
	\bibfield  {author} {\bibinfo {author} {\bibfnamefont {R.~S.}\ \bibnamefont
			{Weis}}\ and\ \bibinfo {author} {\bibfnamefont {T.~K.}\ \bibnamefont
			{Gaylord}},\ }\href {\doibase 10.1007/BF00614817} {\bibfield  {journal}
		{\bibinfo  {journal} {Applied Physics A Solids and Surfaces}\ }\textbf
		{\bibinfo {volume} {37}},\ \bibinfo {pages} {191} (\bibinfo {year}
		{1985})}\BibitemShut {NoStop}%
	\bibitem [{\citenamefont {Bazzan}\ and\ \citenamefont
		{Sada}(2015)}]{Bazzan2015}%
	\BibitemOpen
	\bibfield  {author} {\bibinfo {author} {\bibfnamefont {M.}~\bibnamefont
			{Bazzan}}\ and\ \bibinfo {author} {\bibfnamefont {C.}~\bibnamefont {Sada}},\
	}\href {\doibase 10.1063/1.4931601} {\bibfield  {journal} {\bibinfo
			{journal} {Applied Physics Reviews}\ }\textbf {\bibinfo {volume} {2}},\
		\bibinfo {pages} {040603} (\bibinfo {year} {2015})}\BibitemShut {NoStop}%
	\bibitem [{\citenamefont {Luo}\ \emph {et~al.}(2015)\citenamefont {Luo},
		\citenamefont {Herrmann}, \citenamefont {Krapick}, \citenamefont {Brecht},
		\citenamefont {Ricken}, \citenamefont {Quiring}, \citenamefont {Suche},
		\citenamefont {Sohler},\ and\ \citenamefont {Silberhorn}}]{Luo2015}%
	\BibitemOpen
	\bibfield  {author} {\bibinfo {author} {\bibfnamefont {K.~H.}\ \bibnamefont
			{Luo}}, \bibinfo {author} {\bibfnamefont {H.}~\bibnamefont {Herrmann}},
		\bibinfo {author} {\bibfnamefont {S.}~\bibnamefont {Krapick}}, \bibinfo
		{author} {\bibfnamefont {B.}~\bibnamefont {Brecht}}, \bibinfo {author}
		{\bibfnamefont {R.}~\bibnamefont {Ricken}}, \bibinfo {author} {\bibfnamefont
			{V.}~\bibnamefont {Quiring}}, \bibinfo {author} {\bibfnamefont
			{H.}~\bibnamefont {Suche}}, \bibinfo {author} {\bibfnamefont
			{W.}~\bibnamefont {Sohler}}, \ and\ \bibinfo {author} {\bibfnamefont
			{C.}~\bibnamefont {Silberhorn}},\ }\href {\doibase
		10.1088/1367-2630/17/7/073039} {\bibfield  {journal} {\bibinfo  {journal}
			{New Journal of Physics}\ }\textbf {\bibinfo {volume} {17}} (\bibinfo {year}
		{2015}),\ 10.1088/1367-2630/17/7/073039},\ \Eprint
	{http://arxiv.org/abs/1504.01854} {arXiv:1504.01854} \BibitemShut {NoStop}%
	\bibitem [{\citenamefont {Boes}\ \emph {et~al.}(2018)\citenamefont {Boes},
		\citenamefont {Corcoran}, \citenamefont {Chang}, \citenamefont {Bowers},\
		and\ \citenamefont {Mitchell}}]{Boes2018}%
	\BibitemOpen
	\bibfield  {author} {\bibinfo {author} {\bibfnamefont {A.}~\bibnamefont
			{Boes}}, \bibinfo {author} {\bibfnamefont {B.}~\bibnamefont {Corcoran}},
		\bibinfo {author} {\bibfnamefont {L.}~\bibnamefont {Chang}}, \bibinfo
		{author} {\bibfnamefont {J.}~\bibnamefont {Bowers}}, \ and\ \bibinfo {author}
		{\bibfnamefont {A.}~\bibnamefont {Mitchell}},\ }\href {\doibase
		10.1002/lpor.201700256} {\bibfield  {journal} {\bibinfo  {journal} {Laser
				{\&} Photonics Reviews}\ }\textbf {\bibinfo {volume} {12}},\ \bibinfo {pages}
		{1700256} (\bibinfo {year} {2018})}\BibitemShut {NoStop}%
	\bibitem [{\citenamefont {Rao}\ and\ \citenamefont {Fathpour}(2018)}]{Rao2018}%
	\BibitemOpen
	\bibfield  {author} {\bibinfo {author} {\bibfnamefont {A.}~\bibnamefont
			{Rao}}\ and\ \bibinfo {author} {\bibfnamefont {S.}~\bibnamefont {Fathpour}},\
	}\href {\doibase 10.1109/JSTQE.2017.2779869} {\bibfield  {journal} {\bibinfo
			{journal} {IEEE Journal of Selected Topics in Quantum Electronics}\ }\textbf
		{\bibinfo {volume} {24}},\ \bibinfo {pages} {1} (\bibinfo {year}
		{2018})}\BibitemShut {NoStop}%
	\bibitem [{\citenamefont {Desiatov}\ \emph {et~al.}(2019)\citenamefont
		{Desiatov}, \citenamefont {Shams-Ansari}, \citenamefont {Zhang},
		\citenamefont {Wang},\ and\ \citenamefont {Lon{\v{c}}ar}}]{Desiatov2019}%
	\BibitemOpen
	\bibfield  {author} {\bibinfo {author} {\bibfnamefont {B.}~\bibnamefont
			{Desiatov}}, \bibinfo {author} {\bibfnamefont {A.}~\bibnamefont
			{Shams-Ansari}}, \bibinfo {author} {\bibfnamefont {M.}~\bibnamefont {Zhang}},
		\bibinfo {author} {\bibfnamefont {C.}~\bibnamefont {Wang}}, \ and\ \bibinfo
		{author} {\bibfnamefont {M.}~\bibnamefont {Lon{\v{c}}ar}},\ }\href {\doibase
		10.1364/OPTICA.6.000380} {\bibfield  {journal} {\bibinfo  {journal} {Optica}\
		}\textbf {\bibinfo {volume} {6}},\ \bibinfo {pages} {380} (\bibinfo {year}
		{2019})},\ \Eprint {http://arxiv.org/abs/1902.08217} {arXiv:1902.08217}
	\BibitemShut {NoStop}%
	\bibitem [{\citenamefont {Liang}\ \emph {et~al.}(2017)\citenamefont {Liang},
		\citenamefont {Luo}, \citenamefont {He}, \citenamefont {Jiang},\ and\
		\citenamefont {Lin}}]{Liang2017}%
	\BibitemOpen
	\bibfield  {author} {\bibinfo {author} {\bibfnamefont {H.}~\bibnamefont
			{Liang}}, \bibinfo {author} {\bibfnamefont {R.}~\bibnamefont {Luo}}, \bibinfo
		{author} {\bibfnamefont {Y.}~\bibnamefont {He}}, \bibinfo {author}
		{\bibfnamefont {H.}~\bibnamefont {Jiang}}, \ and\ \bibinfo {author}
		{\bibfnamefont {Q.}~\bibnamefont {Lin}},\ }\href {\doibase
		10.1364/OPTICA.4.001251} {\bibfield  {journal} {\bibinfo  {journal} {Optica}\
		}\textbf {\bibinfo {volume} {4}},\ \bibinfo {pages} {1251} (\bibinfo {year}
		{2017})},\ \Eprint {http://arxiv.org/abs/1706.08904} {arXiv:1706.08904}
	\BibitemShut {NoStop}%
	\bibitem [{\citenamefont {Rao}\ \emph {et~al.}(2017)\citenamefont {Rao},
		\citenamefont {Chiles}, \citenamefont {Khan}, \citenamefont {Toroghi},
		\citenamefont {Malinowski}, \citenamefont {Camacho-Gonz{\'{a}}lez},\ and\
		\citenamefont {Fathpour}}]{Rao2017}%
	\BibitemOpen
	\bibfield  {author} {\bibinfo {author} {\bibfnamefont {A.}~\bibnamefont
			{Rao}}, \bibinfo {author} {\bibfnamefont {J.}~\bibnamefont {Chiles}},
		\bibinfo {author} {\bibfnamefont {S.}~\bibnamefont {Khan}}, \bibinfo {author}
		{\bibfnamefont {S.}~\bibnamefont {Toroghi}}, \bibinfo {author} {\bibfnamefont
			{M.}~\bibnamefont {Malinowski}}, \bibinfo {author} {\bibfnamefont {G.~F.}\
			\bibnamefont {Camacho-Gonz{\'{a}}lez}}, \ and\ \bibinfo {author}
		{\bibfnamefont {S.}~\bibnamefont {Fathpour}},\ }\href {\doibase
		10.1063/1.4978696} {\bibfield  {journal} {\bibinfo  {journal} {Applied
				Physics Letters}\ }\textbf {\bibinfo {volume} {110}},\ \bibinfo {pages}
		{111109} (\bibinfo {year} {2017})},\ \Eprint
	{http://arxiv.org/abs/1610.02111} {arXiv:1610.02111} \BibitemShut {NoStop}%
	\bibitem [{\citenamefont {Cai}, \citenamefont {Wang},\ and\ \citenamefont
		{Hu}(2015)}]{Cai2015a}%
	\BibitemOpen
	\bibfield  {author} {\bibinfo {author} {\bibfnamefont {L.}~\bibnamefont
			{Cai}}, \bibinfo {author} {\bibfnamefont {Y.}~\bibnamefont {Wang}}, \ and\
		\bibinfo {author} {\bibfnamefont {H.}~\bibnamefont {Hu}},\ }\href {\doibase
		10.1364/OL.40.003013} {\bibfield  {journal} {\bibinfo  {journal} {Optics
				Letters}\ }\textbf {\bibinfo {volume} {40}},\ \bibinfo {pages} {3013}
		(\bibinfo {year} {2015})},\ \Eprint {http://arxiv.org/abs/arXiv:1409.6351v1}
	{arXiv:arXiv:1409.6351v1} \BibitemShut {NoStop}%
	\bibitem [{\citenamefont {Cai}, \citenamefont {Kang},\ and\ \citenamefont
		{Hu}(2016)}]{Cai2016a}%
	\BibitemOpen
	\bibfield  {author} {\bibinfo {author} {\bibfnamefont {L.}~\bibnamefont
			{Cai}}, \bibinfo {author} {\bibfnamefont {Y.}~\bibnamefont {Kang}}, \ and\
		\bibinfo {author} {\bibfnamefont {H.}~\bibnamefont {Hu}},\ }\href {\doibase
		10.1364/OE.24.004640} {\bibfield  {journal} {\bibinfo  {journal} {Optics
				Express}\ }\textbf {\bibinfo {volume} {24}},\ \bibinfo {pages} {4640}
		(\bibinfo {year} {2016})}\BibitemShut {NoStop}%
	\bibitem [{\citenamefont {Weigel}\ \emph {et~al.}(2016)\citenamefont {Weigel},
		\citenamefont {Savanier}, \citenamefont {DeRose}, \citenamefont {Pomerene},
		\citenamefont {Starbuck}, \citenamefont {Lentine}, \citenamefont {Stenger},\
		and\ \citenamefont {Mookherjea}}]{Weigel2016}%
	\BibitemOpen
	\bibfield  {author} {\bibinfo {author} {\bibfnamefont {P.~O.}\ \bibnamefont
			{Weigel}}, \bibinfo {author} {\bibfnamefont {M.}~\bibnamefont {Savanier}},
		\bibinfo {author} {\bibfnamefont {C.~T.}\ \bibnamefont {DeRose}}, \bibinfo
		{author} {\bibfnamefont {A.~T.}\ \bibnamefont {Pomerene}}, \bibinfo {author}
		{\bibfnamefont {A.~L.}\ \bibnamefont {Starbuck}}, \bibinfo {author}
		{\bibfnamefont {A.~L.}\ \bibnamefont {Lentine}}, \bibinfo {author}
		{\bibfnamefont {V.}~\bibnamefont {Stenger}}, \ and\ \bibinfo {author}
		{\bibfnamefont {S.}~\bibnamefont {Mookherjea}},\ }\href {\doibase
		10.1038/srep22301} {\bibfield  {journal} {\bibinfo  {journal} {Scientific
				Reports}\ }\textbf {\bibinfo {volume} {6}},\ \bibinfo {pages} {22301}
		(\bibinfo {year} {2016})}\BibitemShut {NoStop}%
	\bibitem [{\citenamefont {Weigel}\ \emph {et~al.}(2018)\citenamefont {Weigel},
		\citenamefont {Zhao}, \citenamefont {Fang}, \citenamefont {Al-Rubaye},
		\citenamefont {Trotter}, \citenamefont {Hood}, \citenamefont {Mudrick},
		\citenamefont {Dallo}, \citenamefont {Pomerene}, \citenamefont {Starbuck},
		\citenamefont {DeRose}, \citenamefont {Lentine}, \citenamefont {Rebeiz},\
		and\ \citenamefont {Mookherjea}}]{Weigel2018e}%
	\BibitemOpen
	\bibfield  {author} {\bibinfo {author} {\bibfnamefont {P.~O.}\ \bibnamefont
			{Weigel}}, \bibinfo {author} {\bibfnamefont {J.}~\bibnamefont {Zhao}},
		\bibinfo {author} {\bibfnamefont {K.}~\bibnamefont {Fang}}, \bibinfo {author}
		{\bibfnamefont {H.}~\bibnamefont {Al-Rubaye}}, \bibinfo {author}
		{\bibfnamefont {D.}~\bibnamefont {Trotter}}, \bibinfo {author} {\bibfnamefont
			{D.}~\bibnamefont {Hood}}, \bibinfo {author} {\bibfnamefont {J.}~\bibnamefont
			{Mudrick}}, \bibinfo {author} {\bibfnamefont {C.}~\bibnamefont {Dallo}},
		\bibinfo {author} {\bibfnamefont {A.~T.}\ \bibnamefont {Pomerene}}, \bibinfo
		{author} {\bibfnamefont {A.~L.}\ \bibnamefont {Starbuck}}, \bibinfo {author}
		{\bibfnamefont {C.~T.}\ \bibnamefont {DeRose}}, \bibinfo {author}
		{\bibfnamefont {A.~L.}\ \bibnamefont {Lentine}}, \bibinfo {author}
		{\bibfnamefont {G.}~\bibnamefont {Rebeiz}}, \ and\ \bibinfo {author}
		{\bibfnamefont {S.}~\bibnamefont {Mookherjea}},\ }\href {\doibase
		10.1364/OE.26.023728} {\bibfield  {journal} {\bibinfo  {journal} {Optics
				Express}\ }\textbf {\bibinfo {volume} {26}},\ \bibinfo {pages} {23728}
		(\bibinfo {year} {2018})},\ \Eprint {http://arxiv.org/abs/1803.10365}
	{arXiv:1803.10365} \BibitemShut {NoStop}%
	\bibitem [{\citenamefont {Volk}, \citenamefont {Gainutdinov},\ and\
		\citenamefont {Zhang}(2017)}]{Volk2017}%
	\BibitemOpen
	\bibfield  {author} {\bibinfo {author} {\bibfnamefont {T.~R.}\ \bibnamefont
			{Volk}}, \bibinfo {author} {\bibfnamefont {R.~V.}\ \bibnamefont
			{Gainutdinov}}, \ and\ \bibinfo {author} {\bibfnamefont {H.~H.}\ \bibnamefont
			{Zhang}},\ }\href {\doibase 10.1063/1.4978857} {\bibfield  {journal}
		{\bibinfo  {journal} {Applied Physics Letters}\ }\textbf {\bibinfo {volume}
			{110}},\ \bibinfo {pages} {132905} (\bibinfo {year} {2017})}\BibitemShut
	{NoStop}%
	\bibitem [{\citenamefont {Shao}\ \emph {et~al.}(2016)\citenamefont {Shao},
		\citenamefont {Bai}, \citenamefont {Cui}, \citenamefont {Li}, \citenamefont
		{Qiu}, \citenamefont {Geng}, \citenamefont {Wu},\ and\ \citenamefont
		{Lu}}]{Shao2016}%
	\BibitemOpen
	\bibfield  {author} {\bibinfo {author} {\bibfnamefont {G.-H.}\ \bibnamefont
			{Shao}}, \bibinfo {author} {\bibfnamefont {Y.-H.}\ \bibnamefont {Bai}},
		\bibinfo {author} {\bibfnamefont {G.-X.}\ \bibnamefont {Cui}}, \bibinfo
		{author} {\bibfnamefont {C.}~\bibnamefont {Li}}, \bibinfo {author}
		{\bibfnamefont {X.-B.}\ \bibnamefont {Qiu}}, \bibinfo {author} {\bibfnamefont
			{D.-Q.}\ \bibnamefont {Geng}}, \bibinfo {author} {\bibfnamefont
			{D.}~\bibnamefont {Wu}}, \ and\ \bibinfo {author} {\bibfnamefont {Y.-Q.}\
			\bibnamefont {Lu}},\ }\href {\doibase 10.1063/1.4959197} {\bibfield
		{journal} {\bibinfo  {journal} {AIP Advances}\ }\textbf {\bibinfo {volume}
			{6}},\ \bibinfo {pages} {075011} (\bibinfo {year} {2016})}\BibitemShut
	{NoStop}%
	\bibitem [{\citenamefont {Aspnes}\ and\ \citenamefont
		{Studna}(1983)}]{Apnes1983}%
	\BibitemOpen
	\bibfield  {author} {\bibinfo {author} {\bibfnamefont {D.~E.}\ \bibnamefont
			{Aspnes}}\ and\ \bibinfo {author} {\bibfnamefont {A.~A.}\ \bibnamefont
			{Studna}},\ }\href {\doibase 10.1103/PhysRevB.27.985} {\bibfield  {journal}
		{\bibinfo  {journal} {Physical Review B}\ }\textbf {\bibinfo {volume} {27}},\
		\bibinfo {pages} {985} (\bibinfo {year} {1983})}\BibitemShut {NoStop}%
	\bibitem [{\citenamefont {Zelmon}, \citenamefont {Small},\ and\ \citenamefont
		{Jundt}(1997)}]{Zelmon2008}%
	\BibitemOpen
	\bibfield  {author} {\bibinfo {author} {\bibfnamefont {D.~E.}\ \bibnamefont
			{Zelmon}}, \bibinfo {author} {\bibfnamefont {D.~L.}\ \bibnamefont {Small}}, \
		and\ \bibinfo {author} {\bibfnamefont {D.}~\bibnamefont {Jundt}},\ }\href
	{\doibase 10.1364/JOSAB.14.003319} {\bibfield  {journal} {\bibinfo  {journal}
			{Journal of the Optical Society of America B}\ }\textbf {\bibinfo {volume}
			{14}},\ \bibinfo {pages} {3319} (\bibinfo {year} {1997})}\BibitemShut
	{NoStop}%
	\bibitem [{\citenamefont {Malitson}(1965)}]{Malitson1965}%
	\BibitemOpen
	\bibfield  {author} {\bibinfo {author} {\bibfnamefont {I.~H.}\ \bibnamefont
			{Malitson}},\ }\href {\doibase 10.1364/JOSA.55.001205} {\bibfield  {journal}
		{\bibinfo  {journal} {Journal of the Optical Society of America}\ }\textbf
		{\bibinfo {volume} {55}},\ \bibinfo {pages} {1205} (\bibinfo {year}
		{1965})}\BibitemShut {NoStop}%
	\bibitem [{\citenamefont {Green}(2008)}]{Green2008}%
	\BibitemOpen
	\bibfield  {author} {\bibinfo {author} {\bibfnamefont {M.~A.}\ \bibnamefont
			{Green}},\ }\href {\doibase 10.1016/j.solmat.2008.06.009} {\bibfield
		{journal} {\bibinfo  {journal} {Solar Energy Materials and Solar Cells}\
		}\textbf {\bibinfo {volume} {92}},\ \bibinfo {pages} {1305} (\bibinfo {year}
		{2008})}\BibitemShut {NoStop}%
	\bibitem [{\citenamefont {Boyd}(2003)}]{Boyd2003}%
	\BibitemOpen
	\bibfield  {author} {\bibinfo {author} {\bibfnamefont {R.~W.}\ \bibnamefont
			{Boyd}},\ }\href@noop {} {\emph {\bibinfo {title} {{Nonlinear Optics}}}}\
	(\bibinfo  {publisher} {Academic Press},\ \bibinfo {address} {London, United
		Kingdom},\ \bibinfo {year} {2003})\BibitemShut {NoStop}%
	\bibitem [{\citenamefont {Sandkuijl}\ \emph {et~al.}(2013)\citenamefont
		{Sandkuijl}, \citenamefont {Tuer}, \citenamefont {Tokarz}, \citenamefont
		{Sipe},\ and\ \citenamefont {Barzda}}]{Sandkuijl2013}%
	\BibitemOpen
	\bibfield  {author} {\bibinfo {author} {\bibfnamefont {D.}~\bibnamefont
			{Sandkuijl}}, \bibinfo {author} {\bibfnamefont {A.~E.}\ \bibnamefont {Tuer}},
		\bibinfo {author} {\bibfnamefont {D.}~\bibnamefont {Tokarz}}, \bibinfo
		{author} {\bibfnamefont {J.~E.}\ \bibnamefont {Sipe}}, \ and\ \bibinfo
		{author} {\bibfnamefont {V.}~\bibnamefont {Barzda}},\ }\href {\doibase
		10.1364/JOSAB.30.000382} {\bibfield  {journal} {\bibinfo  {journal} {Journal
				of the Optical Society of America B}\ }\textbf {\bibinfo {volume} {30}},\
		\bibinfo {pages} {382} (\bibinfo {year} {2013})}\BibitemShut {NoStop}%
	\bibitem [{\citenamefont {Sandkuijl}(2013)}]{Sandkuijl2013a}%
	\BibitemOpen
	\bibfield  {author} {\bibinfo {author} {\bibfnamefont {D.}~\bibnamefont
			{Sandkuijl}},\ }\emph {\bibinfo {title} {{New harmonic generation microscopy
				techniques based on focal volume modelling}}},\ \href@noop {} {Ph.D.
		thesis},\ \bibinfo  {school} {University of Toronto} (\bibinfo {year}
	{2013})\BibitemShut {NoStop}%
	\bibitem [{Note1()}]{Note1}%
	\BibitemOpen
	\bibinfo {note} {D. Sandkuijl, Computational code for second and third
		harmonic generation in layered media with high numerical aperture focusing,
		http://hdl.handle.net/1807/32992}\BibitemShut {NoStop}%
	\bibitem [{\citenamefont {Saito}\ \emph {et~al.}(2008)\citenamefont {Saito},
		\citenamefont {Kobayashi}, \citenamefont {Hiraga}, \citenamefont {Fujita},
		\citenamefont {Kawano}, \citenamefont {Smith}, \citenamefont {Inouye},\ and\
		\citenamefont {Kawata}}]{Saito2008}%
	\BibitemOpen
	\bibfield  {author} {\bibinfo {author} {\bibfnamefont {Y.}~\bibnamefont
			{Saito}}, \bibinfo {author} {\bibfnamefont {M.}~\bibnamefont {Kobayashi}},
		\bibinfo {author} {\bibfnamefont {D.}~\bibnamefont {Hiraga}}, \bibinfo
		{author} {\bibfnamefont {K.}~\bibnamefont {Fujita}}, \bibinfo {author}
		{\bibfnamefont {S.}~\bibnamefont {Kawano}}, \bibinfo {author} {\bibfnamefont
			{N.~I.}\ \bibnamefont {Smith}}, \bibinfo {author} {\bibfnamefont
			{Y.}~\bibnamefont {Inouye}}, \ and\ \bibinfo {author} {\bibfnamefont
			{S.}~\bibnamefont {Kawata}},\ }\href {\doibase 10.1002/jrs.1953} {\bibfield
		{journal} {\bibinfo  {journal} {Journal of Raman Spectroscopy}\ }\textbf
		{\bibinfo {volume} {39}},\ \bibinfo {pages} {1643} (\bibinfo {year}
		{2008})},\ \Eprint {http://arxiv.org/abs/arXiv:1011.1669v3}
	{arXiv:arXiv:1011.1669v3} \BibitemShut {NoStop}%
	\bibitem [{\citenamefont {Jain}\ \emph {et~al.}(2006)\citenamefont {Jain},
		\citenamefont {Lotsberg}, \citenamefont {Stamnes},\ and\ \citenamefont
		{Frette}}]{Jain2006a}%
	\BibitemOpen
	\bibfield  {author} {\bibinfo {author} {\bibfnamefont {M.}~\bibnamefont
			{Jain}}, \bibinfo {author} {\bibfnamefont {J.}~\bibnamefont {Lotsberg}},
		\bibinfo {author} {\bibfnamefont {J.}~\bibnamefont {Stamnes}}, \ and\
		\bibinfo {author} {\bibfnamefont {{\O}.}~\bibnamefont {Frette}},\ }\href
	{\doibase 10.1016/j.optcom.2006.05.047} {\bibfield  {journal} {\bibinfo
			{journal} {Optics Communications}\ }\textbf {\bibinfo {volume} {266}},\
		\bibinfo {pages} {438} (\bibinfo {year} {2006})}\BibitemShut {NoStop}%
	\bibitem [{\citenamefont {Gusachenko}\ and\ \citenamefont
		{Schanne-Klein}(2013)}]{Gusachenko2013}%
	\BibitemOpen
	\bibfield  {author} {\bibinfo {author} {\bibfnamefont {I.}~\bibnamefont
			{Gusachenko}}\ and\ \bibinfo {author} {\bibfnamefont {M.~C.}\ \bibnamefont
			{Schanne-Klein}},\ }\href {\doibase 10.1103/PhysRevA.88.053811} {\bibfield
		{journal} {\bibinfo  {journal} {Physical Review A - Atomic, Molecular, and
				Optical Physics}\ }\textbf {\bibinfo {volume} {88}},\ \bibinfo {pages} {1}
		(\bibinfo {year} {2013})}\BibitemShut {NoStop}%
	\bibitem [{\citenamefont {Nikogosjan}(2005)}]{Nikogosjan2005}%
	\BibitemOpen
	\bibfield  {author} {\bibinfo {author} {\bibfnamefont {D.}~\bibnamefont
			{Nikogosjan}},\ }\href@noop {} {\emph {\bibinfo {title} {{Nonlinear Optical
					Crystals: A Complete Survey}}}}\ (\bibinfo  {publisher} {Springer},\ \bibinfo
	{address} {New York},\ \bibinfo {year} {2005})\BibitemShut {NoStop}%
	\bibitem [{\citenamefont {R{\"{u}}sing}\ \emph {et~al.}(2018)\citenamefont
		{R{\"{u}}sing}, \citenamefont {Neufeld}, \citenamefont {Brockmeier},
		\citenamefont {Eigner}, \citenamefont {Mackwitz}, \citenamefont {Spychala},
		\citenamefont {Silberhorn}, \citenamefont {Schmidt}, \citenamefont {Berth},
		\citenamefont {Zrenner},\ and\ \citenamefont {Sanna}}]{Rusing2018a}%
	\BibitemOpen
	\bibfield  {author} {\bibinfo {author} {\bibfnamefont {M.}~\bibnamefont
			{R{\"{u}}sing}}, \bibinfo {author} {\bibfnamefont {S.}~\bibnamefont
			{Neufeld}}, \bibinfo {author} {\bibfnamefont {J.}~\bibnamefont {Brockmeier}},
		\bibinfo {author} {\bibfnamefont {C.}~\bibnamefont {Eigner}}, \bibinfo
		{author} {\bibfnamefont {P.}~\bibnamefont {Mackwitz}}, \bibinfo {author}
		{\bibfnamefont {K.}~\bibnamefont {Spychala}}, \bibinfo {author}
		{\bibfnamefont {C.}~\bibnamefont {Silberhorn}}, \bibinfo {author}
		{\bibfnamefont {W.~G.}\ \bibnamefont {Schmidt}}, \bibinfo {author}
		{\bibfnamefont {G.}~\bibnamefont {Berth}}, \bibinfo {author} {\bibfnamefont
			{A.}~\bibnamefont {Zrenner}}, \ and\ \bibinfo {author} {\bibfnamefont
			{S.}~\bibnamefont {Sanna}},\ }\href {\doibase
		10.1103/PhysRevMaterials.2.103801} {\bibfield  {journal} {\bibinfo  {journal}
			{Physical Review Materials}\ }\textbf {\bibinfo {volume} {2}},\ \bibinfo
		{pages} {103801} (\bibinfo {year} {2018})}\BibitemShut {NoStop}%
	\bibitem [{\citenamefont {Dierolf}\ and\ \citenamefont
		{Sandmann}(2007)}]{Dierolf2007}%
	\BibitemOpen
	\bibfield  {author} {\bibinfo {author} {\bibfnamefont {V.}~\bibnamefont
			{Dierolf}}\ and\ \bibinfo {author} {\bibfnamefont {C.}~\bibnamefont
			{Sandmann}},\ }\href {\doibase 10.1016/j.jlumin.2006.08.054} {\bibfield
		{journal} {\bibinfo  {journal} {Journal of Luminescence}\ }\textbf {\bibinfo
			{volume} {125}},\ \bibinfo {pages} {67} (\bibinfo {year} {2007})}\BibitemShut
	{NoStop}%
	\bibitem [{\citenamefont {Wittborn}\ \emph {et~al.}(2002)\citenamefont
		{Wittborn}, \citenamefont {Canalias}, \citenamefont {Rao}, \citenamefont
		{Clemens}, \citenamefont {Karlsson},\ and\ \citenamefont
		{Laurell}}]{Wittborn2002}%
	\BibitemOpen
	\bibfield  {author} {\bibinfo {author} {\bibfnamefont {J.}~\bibnamefont
			{Wittborn}}, \bibinfo {author} {\bibfnamefont {C.}~\bibnamefont {Canalias}},
		\bibinfo {author} {\bibfnamefont {K.~V.}\ \bibnamefont {Rao}}, \bibinfo
		{author} {\bibfnamefont {R.}~\bibnamefont {Clemens}}, \bibinfo {author}
		{\bibfnamefont {H.}~\bibnamefont {Karlsson}}, \ and\ \bibinfo {author}
		{\bibfnamefont {F.}~\bibnamefont {Laurell}},\ }\href {\doibase
		10.1063/1.1455700} {\bibfield  {journal} {\bibinfo  {journal} {Applied
				Physics Letters}\ }\textbf {\bibinfo {volume} {80}},\ \bibinfo {pages} {1622}
		(\bibinfo {year} {2002})}\BibitemShut {NoStop}%
	\bibitem [{\citenamefont {Choi}\ \emph {et~al.}(2012)\citenamefont {Choi},
		\citenamefont {Ko}, \citenamefont {Ro},\ and\ \citenamefont {Yu}}]{Choi2012}%
	\BibitemOpen
	\bibfield  {author} {\bibinfo {author} {\bibfnamefont {J.~W.}\ \bibnamefont
			{Choi}}, \bibinfo {author} {\bibfnamefont {D.~K.}\ \bibnamefont {Ko}},
		\bibinfo {author} {\bibfnamefont {J.~H.}\ \bibnamefont {Ro}}, \ and\ \bibinfo
		{author} {\bibfnamefont {N.~E.}\ \bibnamefont {Yu}},\ }\href {\doibase
		10.1080/00150193.2012.741925} {\bibfield  {journal} {\bibinfo  {journal}
			{Ferroelectrics}\ }\textbf {\bibinfo {volume} {439}},\ \bibinfo {pages} {13}
		(\bibinfo {year} {2012})}\BibitemShut {NoStop}%
	\bibitem [{\citenamefont {Gui}\ \emph {et~al.}(2009)\citenamefont {Gui},
		\citenamefont {Hu}, \citenamefont {Garcia-Granda},\ and\ \citenamefont
		{Sohler}}]{Gui2009}%
	\BibitemOpen
	\bibfield  {author} {\bibinfo {author} {\bibfnamefont {L.}~\bibnamefont
			{Gui}}, \bibinfo {author} {\bibfnamefont {H.}~\bibnamefont {Hu}}, \bibinfo
		{author} {\bibfnamefont {M.}~\bibnamefont {Garcia-Granda}}, \ and\ \bibinfo
		{author} {\bibfnamefont {W.}~\bibnamefont {Sohler}},\ }\href {\doibase
		10.1364/OE.17.003923} {\bibfield  {journal} {\bibinfo  {journal} {Optics
				Express}\ }\textbf {\bibinfo {volume} {17}},\ \bibinfo {pages} {3923}
		(\bibinfo {year} {2009})}\BibitemShut {NoStop}%
	\bibitem [{\citenamefont {Scrymgeour}\ \emph {et~al.}(2005)\citenamefont
		{Scrymgeour}, \citenamefont {Gopalan}, \citenamefont {Itagi}, \citenamefont
		{Saxena},\ and\ \citenamefont {Swart}}]{Scrymgeour2005a}%
	\BibitemOpen
	\bibfield  {author} {\bibinfo {author} {\bibfnamefont {D.~A.}\ \bibnamefont
			{Scrymgeour}}, \bibinfo {author} {\bibfnamefont {V.}~\bibnamefont {Gopalan}},
		\bibinfo {author} {\bibfnamefont {A.}~\bibnamefont {Itagi}}, \bibinfo
		{author} {\bibfnamefont {A.}~\bibnamefont {Saxena}}, \ and\ \bibinfo {author}
		{\bibfnamefont {P.~J.}\ \bibnamefont {Swart}},\ }\href {\doibase
		10.1103/PhysRevB.71.184110} {\bibfield  {journal} {\bibinfo  {journal}
			{Physical Review B}\ }\textbf {\bibinfo {volume} {71}},\ \bibinfo {pages}
		{184110} (\bibinfo {year} {2005})},\ \Eprint {http://arxiv.org/abs/0503312}
	{arXiv:0503312 [cond-mat]} \BibitemShut {NoStop}%
	\bibitem [{\citenamefont {Zipfel}, \citenamefont {Williams},\ and\
		\citenamefont {Webb}(2003)}]{Zipfel2003}%
	\BibitemOpen
	\bibfield  {author} {\bibinfo {author} {\bibfnamefont {W.~R.}\ \bibnamefont
			{Zipfel}}, \bibinfo {author} {\bibfnamefont {R.~M.}\ \bibnamefont
			{Williams}}, \ and\ \bibinfo {author} {\bibfnamefont {W.~W.}\ \bibnamefont
			{Webb}},\ }\href {\doibase 10.1038/nbt899} {\bibfield  {journal} {\bibinfo
			{journal} {Nature Biotechnology}\ }\textbf {\bibinfo {volume} {21}},\
		\bibinfo {pages} {1369} (\bibinfo {year} {2003})}\BibitemShut {NoStop}%
	\bibitem [{\citenamefont {Sanna}\ and\ \citenamefont
		{Schmidt}(2017)}]{Sanna2017}%
	\BibitemOpen
	\bibfield  {author} {\bibinfo {author} {\bibfnamefont {S.}~\bibnamefont
			{Sanna}}\ and\ \bibinfo {author} {\bibfnamefont {W.~G.}\ \bibnamefont
			{Schmidt}},\ }\href {\doibase 10.1088/1361-648X/aa818d} {\bibfield  {journal}
		{\bibinfo  {journal} {Journal of Physics: Condensed Matter}\ }\textbf
		{\bibinfo {volume} {29}},\ \bibinfo {pages} {413001} (\bibinfo {year}
		{2017})}\BibitemShut {NoStop}%
\end{thebibliography}
\end{document}